\documentclass[runningheads]{llncs}
\usepackage[T1]{fontenc}
\usepackage{placeins}
\usepackage{graphicx}
\usepackage{xcolor}
\usepackage{comment}
\usepackage{subcaption}
\usepackage{url}
\usepackage{hyperref}
\usepackage{wrapfig}
\usepackage{bm}
\usepackage{siunitx}
\usepackage{amssymb} 
\usepackage{scalerel}
\usepackage{nicefrac}
\usepackage{placeins}

\usepackage[symbol]{footmisc}

\NewDocumentCommand\handshake{}{\scalerel*{\includegraphics{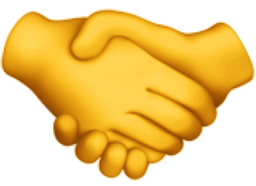}}{\textrm{\textbigcircle}}}

\begin{document}
\title{\textsc{PAKT}: Perspectivized Argumentation Knowledge Graph and Tool for Deliberation Analysis \handshake\\(with Supplementary Materials)}
\titlerunning{Perspectivized Argumentation Knowledge Graph for Deliberation Analysis}

\author{Moritz Plenz$^\ddag$
\inst{1}\orcidID{0009-0001-3316-1359} \and
Philipp Heinisch$^\ddag$
\inst{2}\orcidID{0009-0002-8079-5570} \and
Anette Frank\inst{1}\orcidID{0000-0003-4706-9817} \and
Philipp Cimiano\inst{2}\orcidID{0000-0002-4771-441X}} 
\authorrunning{M. Plenz, P. Heinisch, A. Frank, P. Cimiano}
\institute{
Department of Computational Linguistics, Heidelberg University, Germany\\
\email{\{plenz,frank\}@uni-heidelberg.de}
\and
CITEC, Bielefeld University, Germany\\
\email{\{pheinisch,cimiano\}@techfak.uni-bielefeld.de}
}
\maketitle              %
\renewcommand{\thefootnote}{\fnsymbol{footnote}}
\footnotetext[3]{The authors contributed equally.}
\renewcommand{\thefootnote}{\arabic{footnote}}
\setcounter{footnote}{0}
\begin{abstract} %
Deliberative processes play a vital role in shaping opinions, decisions and policies in our society.  
In contrast to 
persuasive debates, deliberation aims to foster un\-der\-standing of conflicting perspectives among in\-te\-rest\-ed parties. The exchange of arguments in deliberation serves to elu\-ci\-date viewpoints, to raise awareness of conflicting interests, and to finally converge on a resolution. 
To better understand and analyze the underlying 
processes of deliberation, we propose \textsc{PAKT}, a Perspectivized Argumentation Knowledge Graph and Tool. %
The graph structures the argumentative space across diverse topics, where arguments i) are divided into premises and conclusions, ii) are annotated for stances, framings and their underlying values and iii) are connected to background knowledge. 
We show how
to construct \textsc{PAKT} and conduct case studies on the obtained multi\-faceted argumentation graph. Our findings show the analytical potential offered by our framework, highlighting the 
capability to go beyond individual arguments and to reveal structural patterns in the way 
participants and stakeholders argue
in a debate. The overarching goal of our work is to facilitate constructive discourse and informed decision making as a special form of argumentation. 
We offer public access to \textsc{PAKT} and its rich capabilities to support analytics, visualizaton, navigation and efficient search, for diverse forms of 
argumentation.\footnote{GitHub: 
\url{www.github.com/Heidelberg-NLP/PAKT} \\
Website: \url{www.webtentacle1.techfak.uni-bielefeld.de/accept/}}


\end{abstract}

\section{Introduction}
Deliberative processes play a vital role in shaping opinions, decisions, and policies in society. 
Deliberation is the collaborative process of discussing contested issues, to collect and form opinions and guide judgment, in order to find consensus among stakeholders. The key underlying idea is that groups are able to make better decisions regarding societal problems than individuals.\footnote{Cf.\ Habermas, Cohen, Dryzek, Fishkin, see \url{https://tinyurl.com/2p9vsha7}.} 
Deliberation thus can change minds and attitudes, provided that participating individuals are willing to communicate, advocate and to become persuaded with and by others~\cite{reiber:2019}.
Effective deliberation, whether in person or online, incorporates sustained and sound modes of argumentation \cite{falk-etal-2021-predicting} and
can take many forms: from (moderated) discussions to role-playing or formal debates. All these activities aim to explore differing perspectives and should lead to informed and inclusive decisions.

Deliberative theory is concerned with investigating and theorizing about how people discuss and come to conclusions. It has been argued that public debates as available in online debating or discussion fora, or social media platforms such as Reddit, are black boxes, as we have little knowledge about how people argue and what their arguments are based upon~\cite{reiber:2019}. Thus, effective tools are needed to shed light on existing debates to better understand how people argue.

In this work we propose a new framework to support advanced analytics of argumentative discourse, which we apply to analyze deliberative discussions, as a special form of argumentation. At the core of our framework is \textsc{PAKT}, a \textit{\underline{P}erspectivized \underline{A}rgumentation \underline{K}nowledge Graph and \underline{T}ool} that relies on a data model suited to formalize and connect argumentative discussions -- be it interactive dialogues or exchanges in Web fora -- enabling a multi-dimensional analysis of the content of arguments, their underlying perspectives and values, and their connection to different stakeholder groups and to background knowledge. 
\textsc{PAKT} builds on the theory of argumentation by segmenting arguments into premises and conclusions, and focuses on their perspectivization by specifying frames and values which arguments highlight or are based on, and using knowledge graphs to ground arguments in relevant background knowledge.

By going beyond single arguments, PAKT characterizes debates at a structural level, revealing patterns in the way specific groups of stakeholders argue and allowing us to analyze important quality aspects of deliberative discussions. %
Hence, PAKT aids in understanding \textit{how people argue}, including question such as 
i) \textit{Given a debated issue, are (all) relevant argumentative perspectives covered?} 
ii) \textit{Who provided which argument(s)?} and \textit{What are common framings, underlying values and perspectives in presenting them?} and
iii) \textit{How do these perspectives and values differ between pro and con sides, and stakeholder groups?} %

We leverage and refine state-of-the-art argument mining and knowledge graph construction methods to build a rich, perspectivized argumentation knowledge graph, by applying them to debates from \texttt{debate.org} (DDO) as a proof of concept. We show how to analyze this graph in view of its underlying model, and how to answer the above questions by applying \textsc{PAKT} as an analytical tool. 

Our main contributions are:
We i) introduce \textsc{PAKT}, a framework for deliberation analysis that we ii) apply to \texttt{debate.org} as a proof of concept.
We iii) demonstrate how to use it to examine deliberative processes, and iv) offer case studies that leverage PAKT to analyze debates from a deliberative viewpoint.

\section{A Data Model for Perspectivized Argumentation %
} \label{sec:model}

\begin{figure}[t]
    \centering
    \includegraphics[width=.75\linewidth]{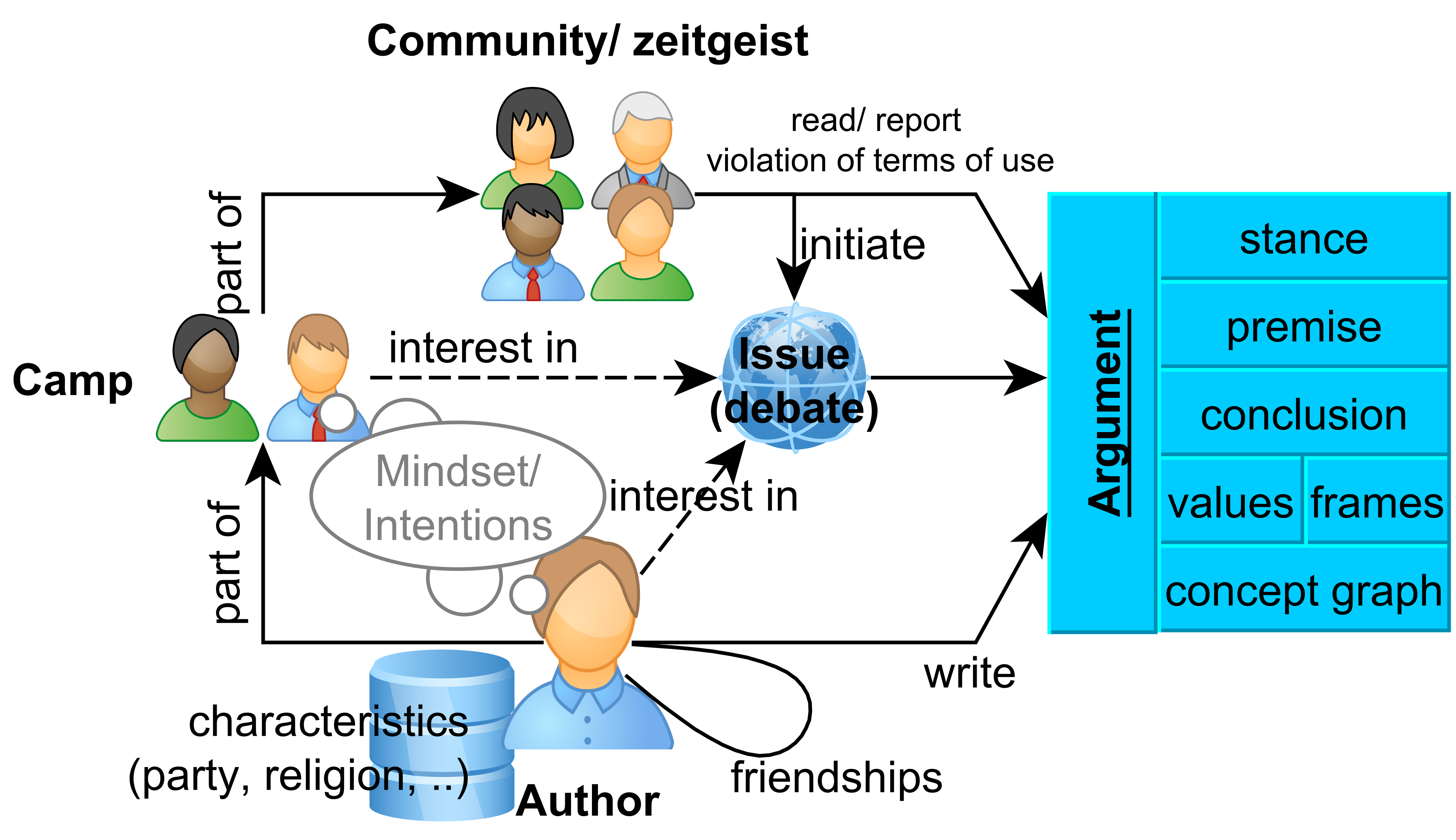}
    \caption{%
PAKT data model consisting of arguments (w/ premises, conclusions, frames, values, stance towards topic and concepts) and authors, camps, zeitgeist
    }
    \label{fig:model}
\end{figure}
Debates in the real world are fundamentally driven by the \textbf{interaction of individuals}. These individuals play various roles in a debate, such as \textit{authors} or members of the \textit{audience}, each bringing unique values, preferred framings and areas of interest into discussions. The \textit{individual characteristics of participants} clearly influence the arguments they formulate and those they  engage with. %

To unravel the complex interplay between individuals and arguments in real-world debates, we present a \textbf{human-centered model} (Fig.~\ref{fig:model}) of a perspectivized argumentation knowledge graph which serves as a structured framework for capturing dynamics in argumentation. Through this formalization, we aim to shed light on the intricacies of framed argumentation, to enhance our understanding of how individuals engage in discussion, and how they can help shaping the quality and outcome of debates, to make them \textit{deliberative}.

\textbf{Authors}, as all individuals, have diverse \textit{beliefs}, \textit{values} and \textit{issues of interest}. 
Individuals who share properties naturally coalesce into \textbf{camps}, which may manifest as formal entities, e.g., political parties, or informal gatherings. Importantly, camps need not adhere to formal memberships, and individuals can participate in multiple camps, even if they hold partially contradictory positions.

By uniting all individuals or camps within a \textbf{community}, we arrive at the concept of the \textit{zeitgeist}—a collective repository of beliefs and norms. 
It governs the relevance and controversy of issues, and thereby shapes the landscape of debates. It also influences the arguments presented within these debates. Arguments that violate the code of conduct, e.g., are typically avoided by authors or moderated out. \textit{Readers}, being part of the community, assess arguments through the lens of the zeitgeist, which can impact their agreement or conviction levels.

Authors, guided by personal convictions or their camps' interests, craft \textbf{arguments} on specific issues. Arguments usually comprise a \textit{premise} and \textit{conclusion}, and reflect a particular \textit{stance} on the issue at hand. Arguments reveal additional information by exposing specific \textit{framings}, \textit{values}, or \textit{concepts} that authors (often deliberately) use to convey their message. Note that these choices can be influenced by the author, their camps, the zeitgeist, or even the audience.

A \textbf{debate} is formed by all arguments on a specific \textbf{issue} put forth by its participants. A good \textit{deliberative} debate should cover all relevant aspects of the issue. This can be achieved by including all \textit{interested parties} and by exploring \textit{(counter-)arguments} of all stances that consider different perspectives and viewpoints of individuals and camps, while ensuring the soundness of each argument.

\section{Constructing \textsc{PAKT}\textsubscript{DDO} from \texttt{debate.org}}
\label{sec:construction}

This section describes, as proof of concept, how we apply PAKT to represent debates from \texttt{debate.org} (\textsc{ddo} for short) and which methods we apply to construct the graph. %
Minor implementation details 
are %
in our %
supplementary materials~\cite{appendix}.

\subsection{Arguments from \texttt{debate.org}} \label{sec:construction:argument}
Fig.~\ref{fig:model} shows two core components of \textsc{PAKT}: i) a \textit{set of arguments} discussing debatable issues and ii) \textit{authors of these arguments}, who can be related to each other. 
While existing argumentative datasets~\cite{stab-gurevych-2017-parsing,ajjour-etal-2019-modeling,kiesel-etal-2023-semeval} do not include author information, a well-known platform that hosts a rich source of arguments along with author profiles is the former debate portal \texttt{debate.org} (DDO).\footnote{The website went offline on 5th of June, 2022. See Fig.~\ref{fig:debateOrgOpinionPoll} for an example screenshot.} This debate portal has been crawled and used in the field of argument mining several times~\cite{wachsmuth:2017f,durmus-cardie-2019-corpus,durmus-cardie-2018-exploring}. To further broaden the extracted data of this portal, we selected 140 controversial issues with at least 25 contributed opinions each, yielding overall 24,646 arguments, where a user profile is available for 7,001 arguments.

\textbf{Stance, premise and conclusion of arguments.}
The DDO portal presents controversial issues as questions that users answer with \emph{yes} (pro) or \emph{no} (con), followed by a \textit{header} and a \textit{statement} (opinion) that explains the answer in detail.
We construct arguments from this data by interpreting the provided statement as the \textit{premise} and automatically generating a \textit{conclusion}. Consider the example: 

\vspace*{3mm}
\begin{tabular}{lp{9.1cm}}
Issue & \phantom{``}Should animal hunting be banned? \\
Stance & \phantom{``}pro \\
Header & \phantom{``}Sport hunting should be banned \\
Statement\hspace{0.5cm} & ``[...] Hunting for fun or sport should be banned. How is~it \phantom{``}fun killing a defenseless animal that's harming no one? [...]'' 
\end{tabular}
\vspace*{3mm}

\textbf{Conclusion generation.} 
Since conclusions are not given in the DDO data, we construct conclusions automatically. For this we apply ChatGPT in a few-shot setting, showing it three examples consisting of i) the question, ii) stance, iii) header, and iv) a manually created conclusion. For our example, the generated conclusion is ``\textit{Sport hunting should be banned in order to protect animals.}'' The complete prompt is shown in our %
supplementary materials~\cite{appendix}.

\subsection{Characterizing arguments for perspectivized argumentation} \label{sec:construction:characterizing}
We enrich arguments with automatically inferred frames, values and concept graphs to enable easy analysis and filtering in PAKT. 

\textbf{Frames.}
To represent specific viewpoints, perspectives, or aspects from which an argument is made, we adopt the notion of ``frames.'' 
While one line of research tailors frame sets to each issue separately, yielding \textit{issue-specific} frame sets~\cite{ajjour-etal-2019-modeling,schiller-etal-2021-aspect,ruckdeschel-wiedemann-2022-boundary}, we aim to generalize frames across diverse issues. We therefore apply the MediaFrames-Set~\cite{boydstun2014}, a \textit{generic} frame set consisting of 15 classes that are applicable across many issues and topics. 

To apply these frames to arguments from DDO, we fine-tune a range of classifiers on a comprehensive training dataset of more than 10,000 newspaper articles that discuss immigration, same-sex marriage, and marijuana, containing 146,001 text spans labeled with a single MediaFrame-class per annotator~\cite{card-etal-2015-media}. To apply this dataset to our argumentative domain, we broaden the annotated spans to sentence level~\cite{HeinischCimiano+2021+59+72}. Since an argument can address more than a single frame~\cite{reimers-etal-2019-classification}, we design the argument-frame classification task as a multi-label problem by combining all annotations for a sentence into a frame target set. To introduce additional samples with more comprehensive text and target frame sets, we merge existing samples pairwise by combining their text and unifying their target frame set.
As processing architecture, we apply different architectures~\cite{heinisch-etal-2023-accept}, and determine LLMs (RoBERTa~\cite{liu2019roberta}\footnote{For further studies in this paper, we apply the checkpoint \url{https://huggingface.co/pheinisch/MediaFrame-Roberta-recall}}) as the best-performing ones.

\textbf{Human Values.}
Since we aim to analyze arguments not as standalone text, but as text written by individuals with intentions and goals, it is also important to analyze the human values~\cite{xie-etal-2019-text,alshomary-etal-2022-moral,kobbe-etal-2020-exploring,kiesel-etal-2022-identifying} underlying a given argument, to infer the authors' beliefs, desirable qualities, and general action paradigms~\cite{kiesel-etal-2022-identifying}. %
The shared task ``SemEval 2023 Task 4: ValueEval''~\cite{kiesel-etal-2023-semeval} popularized the Schwartz' value continuum~\cite{schwartz-1994-universal}. This is a hierarchical system with four higher-order categories: ``Openness to change'', ``Self-enhancement'', ``Conversation'', and ``Self-transcendence''. At the second level, these categories are refined into 12 categories, including ``Self-direction'', ``Power'', ``Security'', or ``Universalism''. 
To reduce the complexity of the value classification task, we follow Kiesel \textit{et al.}~\cite{kiesel-etal-2023-semeval} in not using the finest granularity of Schwartz' value continuum, but rather the second-smallest %
level containing 20 classes. For predicting value classes for an argument, we rely on a fine-tuned ensemble of three LLMs %
published by the winning team~\cite{schroter-etal-2023-adam} of the shared task. %

\textbf{Concepts.} Humans possess rich commonsense knowledge that allows them to communicate efficiently, by leaving information implicit that can be easily inferred in communication by other humans. Also in argumentation, it is often left implicit how a conclusion follows from a given premise. To uncover which concepts are covered in a given argument -- either explicitly or implicitly -- we link arguments to ConceptNet~\cite{speer-etal-2017-conceptnet}, a popular commonsense knowledge graph. 

To do this we rely on~\cite{plenz-etal-2023-similarity} to extract subgraphs from ConceptNet: We
split the premise into individual sentences (cf.~\cite{heinisch-etal-2023-accept}), %
then, for each sentence in the premise and for the conclusion, we extract relevant ConceptNet concepts. 
These concepts represent explicit mentions in the premise and conclusion, but not implicit connections. 
Hence, we connect the extracted concepts with weighted shortest paths extracted from ConceptNet. 
These paths reveal how the conclusion follows from the premise, along with other potential implicit connections \cite{%
plenz-etal-2023-similarity}.

\subsection{Authors and camps}\label{sec:construction:authors}

In DDO, authors could choose to reveal their user profile when posting an argument. %
To model stakeholder groups, we group users into camps using their user profiles. 
The profiles state distinct categories for traits such as \textit{gender}, \textit{ideology}, \textit{religion}, \textit{income}, or \textit{education}. %
Users could also fill free-text fields about, e.g.,  personal beliefs or quotes. Users control which parts of their profiles are public, so the amount of available data differs for each user. %
To obtain camps, we cluster the stated categories in coarse groups, e.g. \textit{left}, \textit{right} and \textit{unknown} for ideology.

\subsection{Implementation and Tools for Building and Using \textsc{PAKT}}
PAKT is designed to aid in future argumentative analysis, so we make it publicly available in several forms. Our website\footnote{\url{https://webtentacle1.techfak.uni-bielefeld.de/accept}} provides a comprehensive overview of issues in PAKT\textsubscript{DDO} in a search interface. 
To enable richer analysis we also make PAKT\textsubscript{DDO} available as a Neo4J\footnote{\url{https://neo4j.com}} graph database that loosely follows the structure shown in Fig.~\ref{fig:model}. Neo4J databases can be queried with \textit{Cypher}, a powerful, yet easy-to-learn querying language similar to SQL, but that supports queries on graphs. Issues, users, arguments, and other entities can efficiently be searched for and filtered in our database. A detailed description on how to utilize our database can be found at \url{www.github.com/Heidelberg-NLP/PAKT}. %

\subsection{Preliminary Evaluation}

To provide a preliminary evaluation of the quality of the PAKT\textsubscript{DDO} graph,
we manually labeled 99 arguments on the issue ``\textit{Should animal hunting be banned?}'' that will be used in our case study (\S\ref{sec:case-study:animal-hunting}). We evaluate the quality of generated conclusions and annotated labels (frames and values), as well as retrieved supporting and counter arguments. %
Each annotation sample includes the stance, the header, and the full statement (premise). 
For each argument, three annotators provided judgments on five questions\footnote{the labels were aggregated using the majority vote}: (i)~\textit{Conclusion quality} (rating the appropriateness of the conclusion generated by ChatGPT):
94/99 conclusions are labeled as appropriate;
(ii)~\textit{Frame identification} (identifying all emphasized aspects): the predictions yield 0.40 micro-F1; (iii)~\textit{Human value detection} (detecting all values encouraged by the argument): again the predictions yield 0.40 micro-F1; (iv)~\textit{Similarity rating} (given two further arguments, rating whether and which argument is more similar): similarity predictions for arguments with the same stance obtained with S$^3$BERT~\cite{opitz-frank-2022-sbert} correlate with annotator judgments with an accuracy of 42\%; (v)~\textit{Counter rating} (given two further arguments, rating whether and which arguments attack the given argument more): the similarity predictions for arguments with the opposite stance obtained from S$^3$BERT~\cite{opitz-frank-2022-sbert} correlate with an accuracy of 40\%. For detailed analysis of the manual study including IAA see our %
supplementary materials~\cite{appendix}.

\section{Analytics applied to \textsc{PAKT}\textsubscript{\textsc{DDO}}}\label{sec:analyse-general}

In this section we analyse PAKT\textsubscript{DDO} at a global level to discover general trends in our data, by aggregating information across all represented issues.

\begin{figure}[t]
    \centering
    \includegraphics[width=1.0\linewidth]{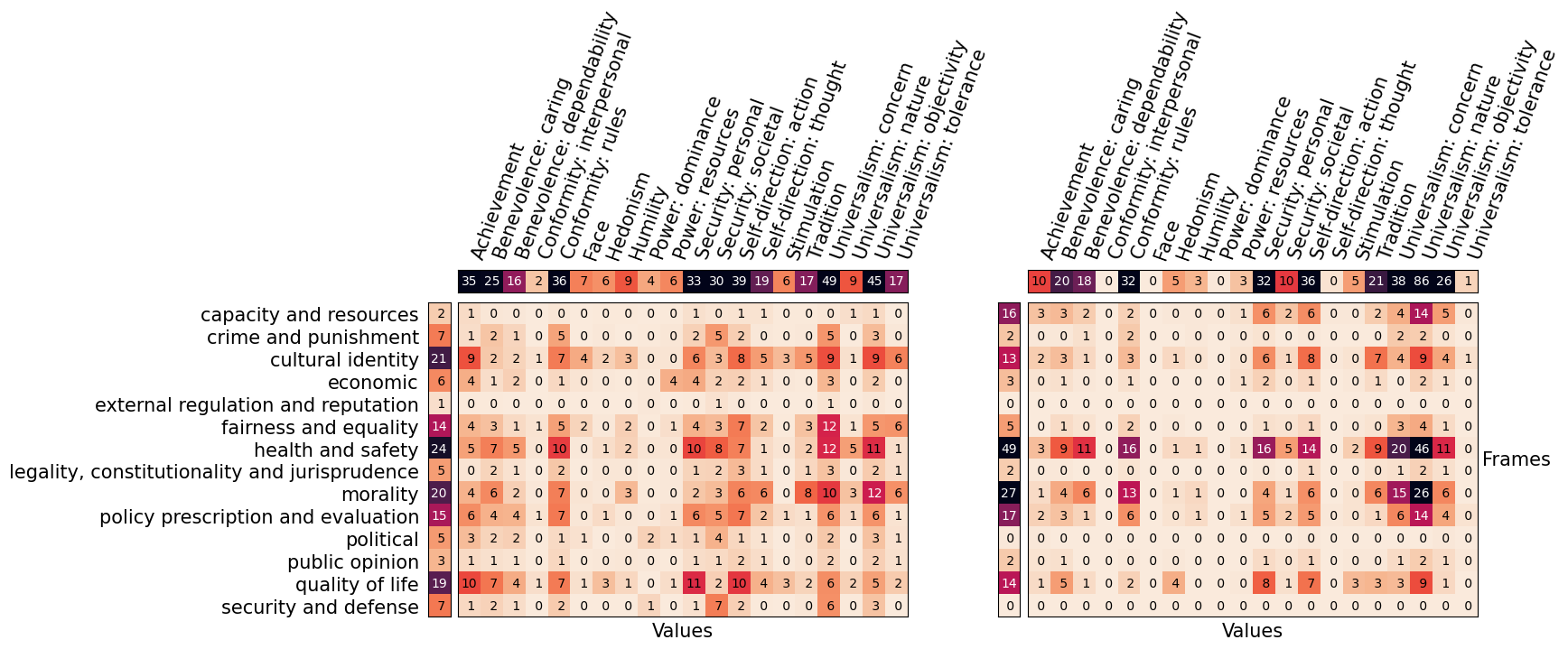}
    \caption{Correlation between frames and values. Left plot is across all topics, right plot is for the issue \textit{Should animal hunting be banned?} Arguments labeled with more than one frame/value are counted multiple times. Numbers are percentages. %
    }
    \label{fig:frames_values_correlation}
\end{figure}

\textbf{Frames and Values.}
Fig. \ref{fig:frames_values_correlation} (left) shows the distribution of frames and human values across all arguments from all issues. The frames \textit{health and safety}, \textit{cultural identity}, \textit{morality} and \textit{quality of life} are the most frequent, each occurring in almost \SI{20}{\percent} of all arguments. The most common values are \textit{concern} (\SI{49}{\%}) and \textit{objectivity} (\SI{45}{\%}). 
We further observe that some frames occur frequently with certain values and vice versa. The \textit{fairness and equality} frame, e.g., occurs six out of seven times in combination with the value \textit{concern}. 

\textbf{Concepts.}
For our analysis in this paper, we consider the ratio of arguments that mention a certain concept. To avoid biases due to the structural properties of ConceptNet (e.g. some concepts are better connected and hence occur more often), we report these ratios relative to the ratio computed over all arguments in PAKT\textsubscript{DDO}. E.g., when reporting the concept ratios for a specific frame, we report the ratio relative to the ratio computed over all arguments that we subtract from the former, i.e., $\frac{N_{fc}}{N_f} - \frac{N_c}{N}$, where $N$ is the number of arguments with a specific frame $f$ or concept $c$. When comparing two subsets of PAKT\textsubscript{DDO} -- for example pro and con on a certain topic -- we instead normalize by the complementary subset to obtain more specific concepts. 

When linking arguments to commonsense background knowledge we see that the most frequent concepts are \textit{Person} and \textit{People}, indicating that most debates are -- as expected -- human-centered. Other commonly occurring concepts are \textit{US}, \textit{Legal}, \textit{War}, or \textit{School} which reflect the categories and context that our issues stem from. These concepts are also frequently used in contemporary debates, which indicates that issues in PAKT\textsubscript{DDO} are representative for general debates. 

Our analysis also reveals concepts that are specific to certain frames and values. For example, the concepts \textit{religion}, \textit{god}, \textit{person}, \textit{biology}, \textit{human} and \textit{christianity} occur between 10 and 24 percentage points (\si{pp}) more often in arguments bearing the \textit{morality} frame, compared to all arguments across all frames.
Similarly, for the value \textit{nature}, the most common concepts are \textit{animals}, \textit{animal}, \textit{zoo}, \textit{kept in zoos}, \textit{killing} and \textit{water}, which occur between 12 and \SI{39}{pp} more often than in all arguments.

\begin{figure}
    \begin{subfigure}{1\textwidth}
        \includegraphics[width=1\linewidth]{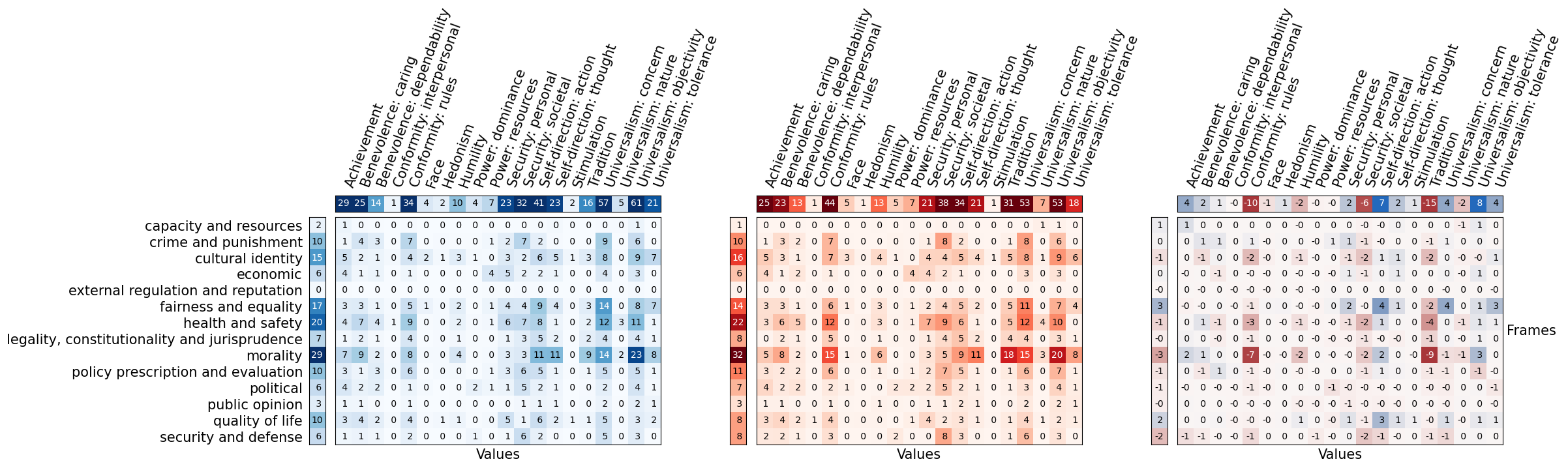}
        \caption{Frames and values across all issues separated by \textbf{author ideology}. \\Left: \textit{left wing}; Middle: \textit{right wing}}
        \label{fig:frames_values_correlation_by_ideology}
    \end{subfigure}
    \begin{subfigure}{1\textwidth}
        \includegraphics[width=\linewidth]{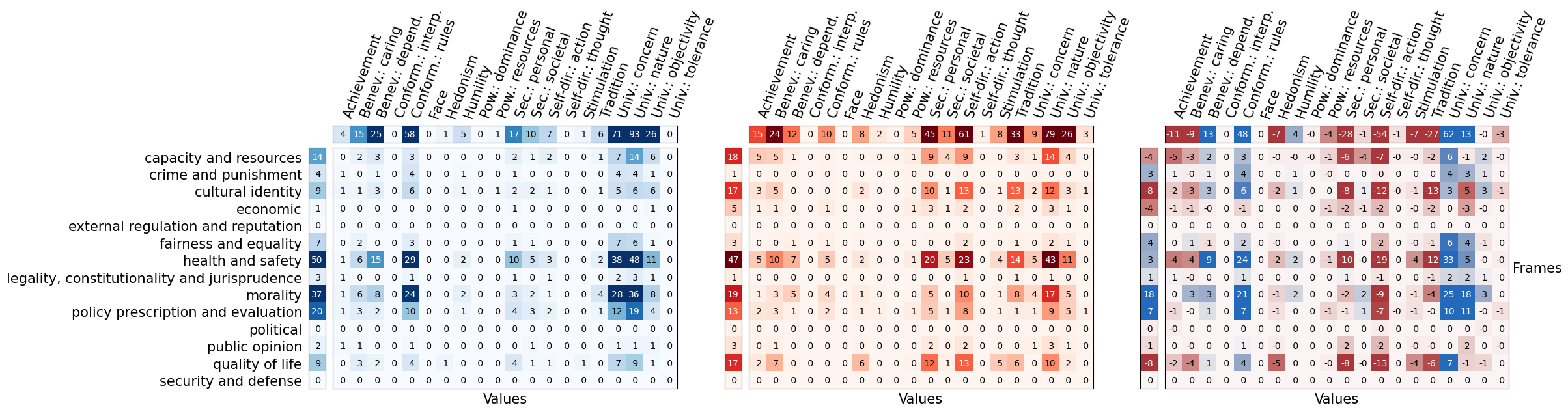}
        \caption{Frames and values for \textit{Should animal hunting be banned?} separated by \textbf{stance}. \\Left: \textit{pro}; Middle: \textit{con}}
        \label{fig:frame_values_hunting_by_stance}
    \end{subfigure}
    \begin{subfigure}{1\textwidth}
        \includegraphics[width=1.0\linewidth]{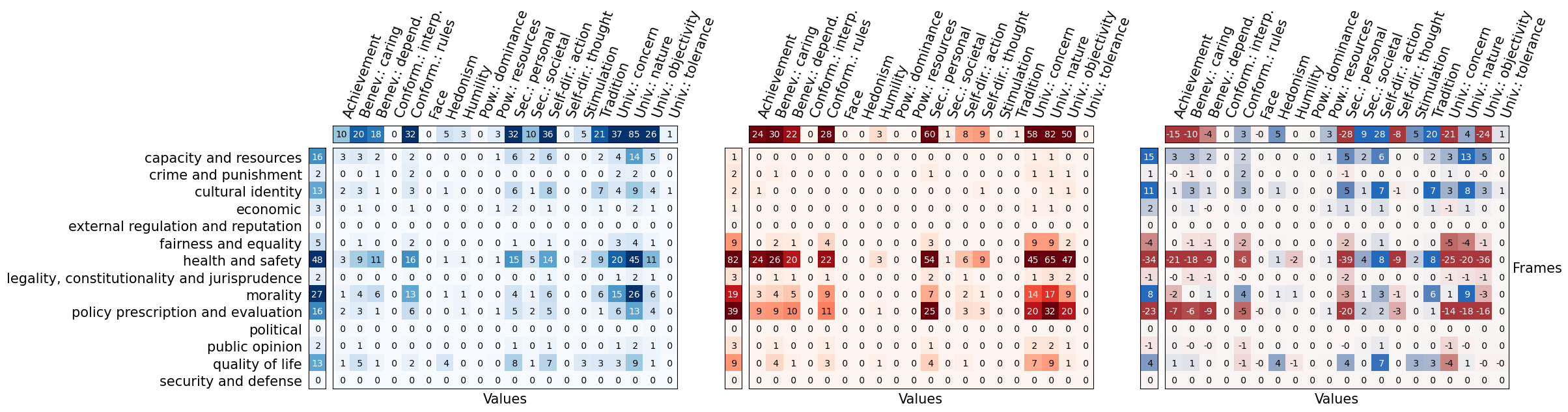}
        \caption{Frames and values for \textbf{different issues}. \\Left: \textit{Should animal hunting be banned?}; Middle: \textit{Should animal testing be banned?}}
        \label{fig:frame_values_different_topics}
    \end{subfigure}
    \begin{subfigure}{1\textwidth}
        \includegraphics[width=1.0\linewidth]{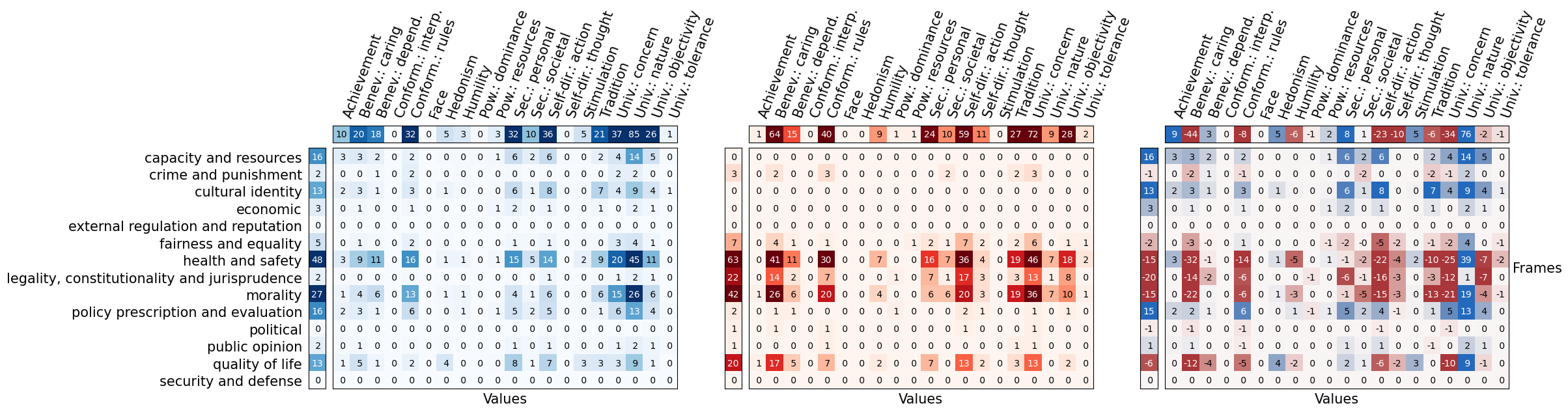}
        \caption{Frames and values for \textbf{different issues}. \\Left: \textit{Should animal hunting be banned?}; Middle: \textit{Should Abortion be illegal in America?}}
        \label{fig:frame_values_huting-vs-abortion}
    \end{subfigure}

    \caption{Comparison between frame and value matrices. 
    The left and middle plots show distributions in percent, and the right plots show their differences in percentage points (pp).}
    \label{fig:frame_values_differences}
\end{figure}

\begin{figure}
    \begin{subfigure}{1\textwidth}
        \includegraphics[width=1\linewidth]{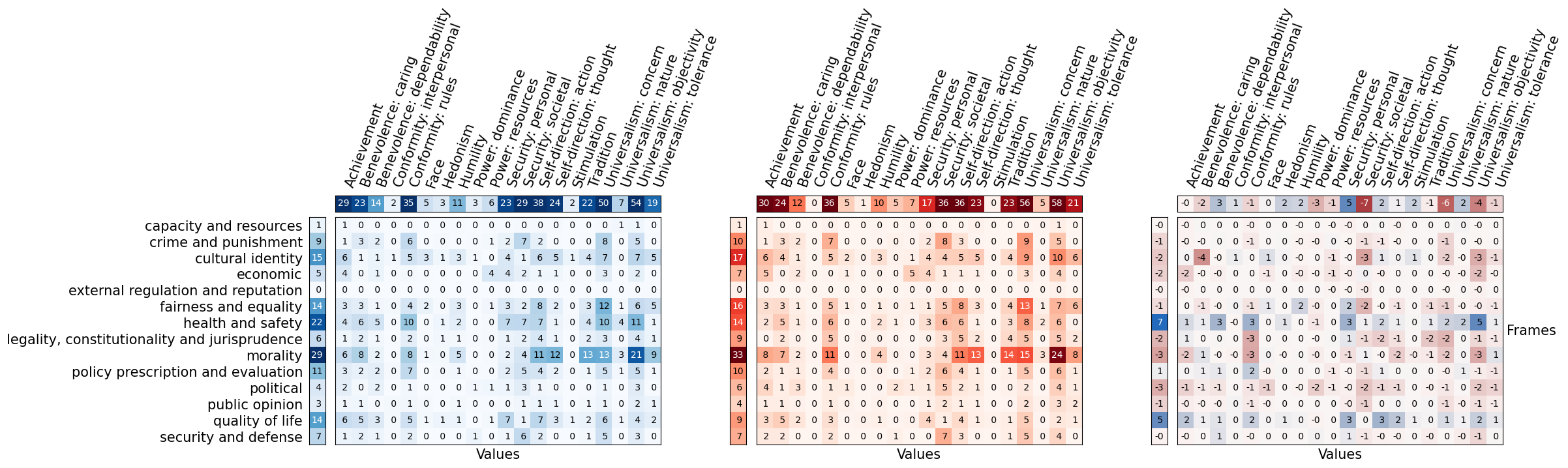}
        \caption{Separated by education. Left: \textit{lower education}; Middle: \textit{higher education}}
    \end{subfigure}
    \begin{subfigure}{1\textwidth}
        \includegraphics[width=1\linewidth]{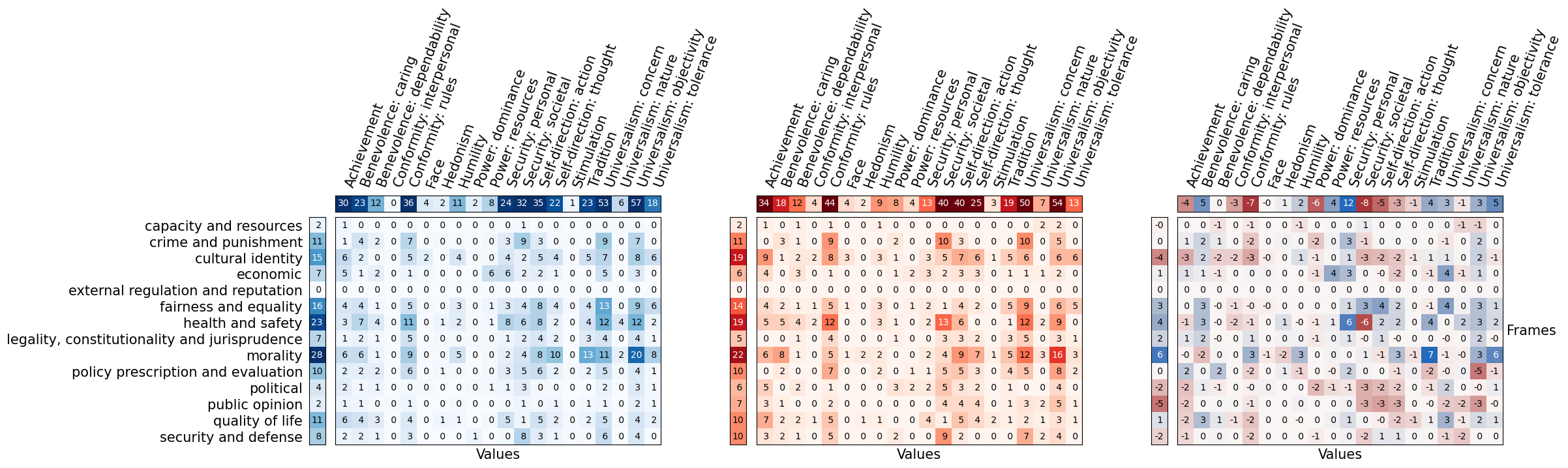}
        \caption{Separated by income. Left: \textit{low income}; Middle: \textit{high income}}
    \end{subfigure}
    \begin{subfigure}{1\textwidth}
        \includegraphics[width=1\linewidth]{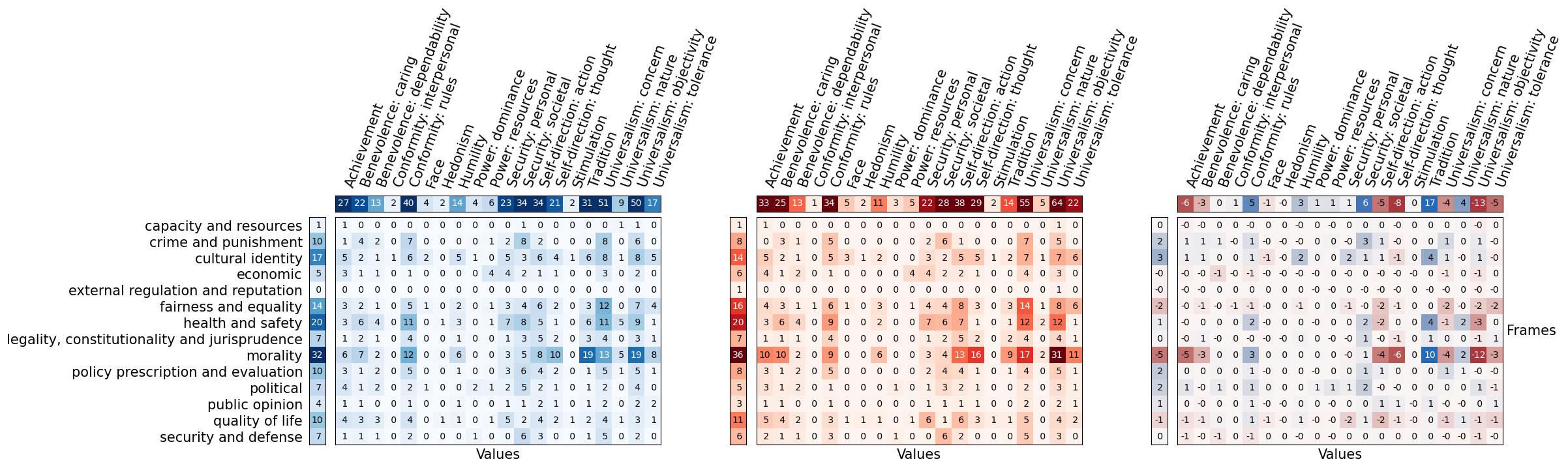}
        \caption{Separated by religion. Left: \textit{yes (i.e. author is religious)}; Middle: \textit{no (i.e. author is not religious})}
    \end{subfigure}
    \begin{subfigure}{1\textwidth}
        \includegraphics[width=1\linewidth]{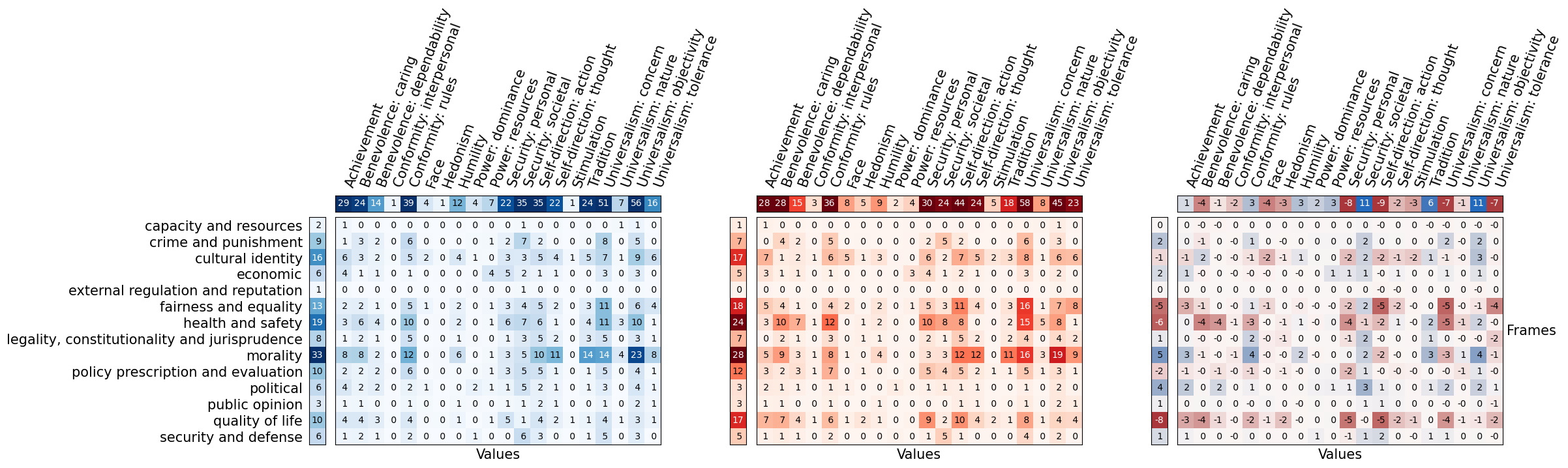}
        \caption{Separated by gender. Left: \textit{male}; Middle: \textit{female}}
    \end{subfigure}

    \caption{Comparison between frame and value matrices. 
    The left and middle plots show distributions in percent, and the right plots show their differences in percentage points (pp). All subfigures are aggregated across all issues.}
    \label{fig:app:frame_value-global_by_stakeholder}
\end{figure}

\textbf{Camps.}
\textsc{PAKT}\textsubscript{\textsc{DDO}}
includes author information 
that users have decided to provide for themselves. Using this information, we can group users (i.e. the authors of arguments) into camps along several dimensions, as described in \S\ref{sec:construction:authors}. 
This allows us to compare which frames and values are preferred by which camps. Fig.~\ref{fig:frames_values_correlation_by_ideology} shows these distribution for authors of different ideology. In comparison, left-winged authors prefer the \textit{objectivity} and \textit{self-direction: action} values, while right-winged authors consider the values \textit{tradition} and \textit{conformity: rules} more. For frames, the difference between the camps is relatively small, indicating that one's ideology is more value-driven. %
Fig.~\ref{fig:app:frame_value-global_by_stakeholder} shows the distributions for other camps, where we observe stronger effects for frames. 

\begin{figure}
    \centering
    \includegraphics[width=0.5\linewidth]{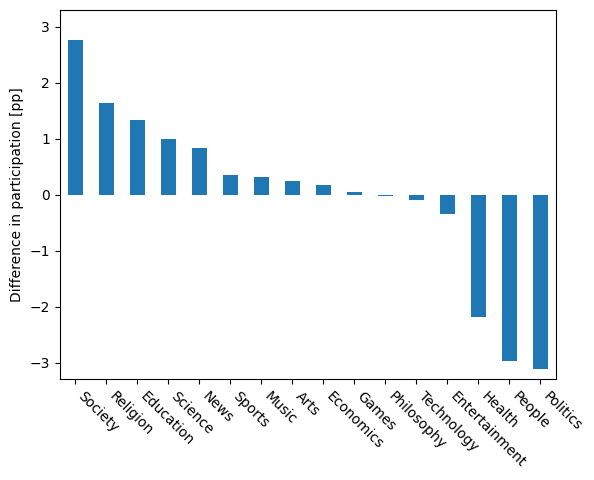}
    \caption{%
    Difference in relative participation between left and right winged authors. 
    }
    \label{fig:app:topic_participation_by_ideology}
\end{figure}

However, since different issues have different relevance for single frames and values, we check whether different distributions of frames and values are caused by different issue participation dependent on the camp. Here, our analysis shows that authors from different camps engage in issues from similar categories, with participation rates differing by at most $\sim$\SI{3}{pp} for ideology (cf. Fig.~\ref{fig:app:topic_participation_by_ideology}), showing that different camps prefer different frames and values while debating on the same issues. 

\section{Case Studies} \label{sec:case_study}

\subsection{Should animal hunting be banned?}\label{sec:case-study:animal-hunting}

For deeper analysis we examine
one specific issue, namely \textit{Should animal hunting be banned?} PAKT\textsubscript{DDO} contains 409 arguments on this issue, with a relatively even parity ($\sim$\SI{46}{\%} pro and \SI{54}{\%} con). 

\begin{figure}
    \center
    \includegraphics[width=0.5\linewidth]{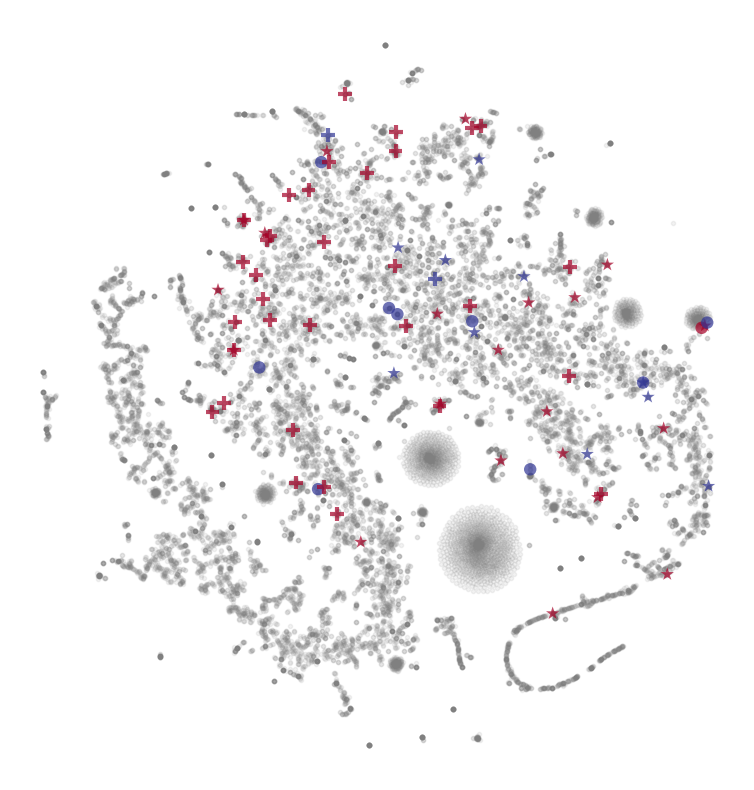}
    \caption{T-SNE embedding of the spectral embeddings of the largest connected component of the friendship network of DDO. Users replying to \textit{Should animal hunting be banned?} ($\star$), \textit{Should animal testing be banned?} ($\bullet$) or \textit{Should humans stop eating animals and become vegetarians?} ($\bm+$) are marked in blue (pro) or red (con). We see that camps are embedded consistently across similar issues.}
    \label{fig:user_network}
\end{figure}

\textbf{Camps.}
Our notion of camps used in \S\ref{sec:analyse-general} requires user information, which is scarce at the level of individual issues. For example, for ideology only 17 contributing authors %
provided user information. Therefore, for the given issue we consider people in favor and against banning animal hunting as distinct camps. Separating authors into camps by their stance actually does reflect the friendship network between authors on DDO, as shown in Fig.~\ref{fig:user_network}. 

\textbf{Frames and Values.} Fig.~\ref{fig:frames_values_correlation} (right) shows the frames and values for this issue. \SI{86}{\%} of arguments address the \textit{nature} value, which is directly linked to the issue. Other frequent values occurring in more than \SI{30}{\%} of arguments are \textit{universalism: concern}, \textit{self-direction: action}, \textit{conformity: rules} and \textit{security: personal}. The most frequent frames are \textit{health and safety} and \textit{morality}. 

To better understand how and why these frames and values arise, we look at how they differ between stances (Fig.~\ref{fig:frame_values_hunting_by_stance}). Firstly, we note that the most frequently occurring frames and values are common for both stances. However, manual inspection of these arguments reveals that these frames and values are interpreted in different ways. For example, on the pro side the \textit{nature} value often refers to species or entire ecosystems being endangered, and that humans should not diminish them even more. By contrast, on the con side, a common interpretation of nature protection is that balance needs to be maintained by hunting over-populating species such as deer. Identifying such shared values with different interpretations can aid in finding common ground and ultimately satisfying compromises. %
Here, a possible compromise could be to ban the hunting of endangered species, but to allow sustainable hunting of certain species. 

However, a value or frame can also predominantly be used by a certain stance. The value \textit{universalism: concern} expresses that all people and animals 
deserve equality, justice, and protection. \SI{71}{\%} of all pro arguments support this value, while only \SI{9}{\%} of all con arguments support it. On the pro side, this value means that we shouldn't hunt animals, as we also would not hunt humans. %
Authors on the con side addressing this value argue that hunted animals have better lives than farmed animals. %
Again, the difference lies in the interpretation.

\textbf{Concepts.} 
For our target issue, we obtain concepts revolving around animals, hunting, killing, and food. Again, we compare pro and con arguments to each other: The most prominent pro-concepts are \textit{killing animal}, \textit{killing}, \textit{bullet}, \textit{animals}, \textit{evil} and \textit{stabbing to death}. On the other hand, the most frequently occurring con-concepts are \textit{getting food}, \textit{fishing}, \textit{eat}, \textit{going fishing}, \textit{meat} and \textit{food}. This highlights the different foci regarding hunting: people in favor of banning hunting emphasize the aspect of killing during hunting, while people who oppose a ban on hunting emphasize the usage of dead animals for food. Hence, the concepts can be seen as issue-specific framings used by the pro and con sides. 

\subsection{Comparison to other issues}
An important aspect of opinion-making, and hence of deliberation, is to learn from similar debates. %
Similar issues can be identified with standard similarity prediction methods like SBERT \cite{reimers-gurevych-2019-sentence,opitz-frank-2022-sbert}, which is already integrated in PAKT. 

\textbf{Frames and Values.} Beyond the similarity of the content of arguments, we may %
be interested in more abstract relations between issues -- for example, we may want to investigate issues with similar frame and value distributions. To detect such issues, we compute the Frobenius norm of the difference between frame-value matrices (cf. Fig.~\ref{fig:frame_values_differences}) of different issues. A small Frobenius norm indicates a similar distribution of emphasized frames and values between the issues.
For animal hunting, the five most similar issues revolve around animals: 
``\textit{Should the United States ban the slaughter of horses for meat?}'', 
``\textit{Should humans stop eating animals and become vegetarians?}'', 
``\textit{Should animals be kept in zoos?}'', 
``\textit{Should we keep animals in zoos?}'' and 
``\textit{Should animal testing be banned?}'' 
The next five most similar issues are 
``\textit{Should cigarette smoking be banned?}'', 
``\textit{Should Abortion be illegal in America?}'', 
``\textit{Pro-life (yes) vs. pro-choice (no)?}'', 
``\textit{Should abortion be illegal?}'' and
``\textit{Does human life begin at conception?}''. 
Four of them are about abortion, which shows that animal rights and abortion evoke similar frames and values (see Fig.~\ref{fig:frame_values_huting-vs-abortion}), perhaps because both issues concern individuals who are unable to defend their own rights. 

In the following we take a closer look at similarities and differences between the issues ``\textit{Should animal hunting be banned?}'' and ``\textit{Should animal testing be banned?}'' We chose these issues, as they seem similar at first glance, but reveal intriguing differences upon closer inspection. Moreover, Fig.~\ref{fig:user_network} shows they have comparable camps. 
As expected, they mostly highlight the same frames and values (Fig.~\ref{fig:frame_values_different_topics}). But there are also notable differences: In \textit{animal testing}, the \textit{health and safety} frame is expressed more often, while \textit{capacity and resources} and \textit{cultural identity} frames are rare. %

Arguments using a \textit{health and safety} frame for a ban on \textit{animal hunting} or \textit{testing} often refer to the health and safety of animals, and to the health and safety of humans when arguing against a ban. 
Yet, the issues raised for the health and safety of humans are not the same in arguments against a ban: 
for animal hunting, a common argument is that humans need meat for nutrition, 
which hunting helps to ensure. For animal testing the health and safety aspect often revolves around animal tests being necessary to make medicine safe for humans. This difference has also very different implications for deliberation. Concerning animal hunting, one could argue that meat for nutrition can be provided by farmed animals, or can be substituted in vegetarian diet. Finding alternatives for animal testing is more difficult and hence, needs to be addressed differently. %

\textbf{Concepts.} Naturally, similar issues share similar concepts, for instance, \textit{animals} in our example, while others are more distinct, e.g., \textit{getting food} for hunting or \textit{scientists} for animal testing. Such differences are often issue-specific and more fine-grained than differences in frames and values, as discussed above. Hence, a deeper analysis of concepts and content can help elucidate potential
differences behind
shared frames and values, which can be important for deliberation.

\subsection{Argument level}\label{sec:case_study:argument-level}
So far, our analysis focused on entire debates, or even collections of debates, to
analyze structural properties, such as similarities and differences among debates.
Yet, PAKT also supports analysis at the level of individual arguments
to enable in-depth analysis. For each argument, PAKT includes abstractions to frames, values, and concepts which is what we mostly used in our analysis so far. 

Beyond this, PAKT allows us to compare and relate arguments based on their content. We can do this by estimating the similarity between arguments, using either S$^3$BERT~\cite{opitz-frank-2022-sbert} or the concept overlap as another interpretable method~\cite{appendix}. 

With the computed similarities, it is almost trivial to retrieve supporting arguments (most similar among the same stance) or counterarguments (most similar but opposing stance)~\cite{wachsmuth-etal-2018-retrieval,shi-etal-2023-revisiting}. More complex argument retrieval is also easy and efficient. For example, to answer the question ``\emph{How would someone argue who wants to make a similar argument but from the perspective of value \emph{x} instead of value \emph{y}?},'' one can use the following query which runs in $\sim$\SI{5}{ms}: \\
\hspace*{5mm}\texttt{MATCH (:argument \{id: \$query\_id\})-[r:SIMILARITY]-(a:argument)\\ 
\hspace*{5mm}WHERE x in a.value AND not y in a.value \\
\hspace*{5mm}RETURN a ORDER BY r.similarity DESC}

\section{Related Work}

A number of approaches have been developed with the goal of analyzing deliberative debates.

Gold \textit{et al.}~\cite{Gold2017} propose an interactive analytical framework that combines linguistic and visual analytics to analyze the quality of deliberative communication
automatically. Deliberative quality is seen as a latent unobserved variable that manifests itself in a number of observable measures and is mainly quantified based on linguistic cues and topical structure. The degree of deliberation
is measured in four dimensions: i) \textit{Participation} considers whether proponents are treated equally, i.e., whether 
all stakeholders are heard; 
ii) \textit{Mutual Respect}
is indicated by  
linguistic markers and patterns of turn-taking; iii) \textit{Argumentation and Justification} aims to ensure that arguments are 
properly justified and refer to agreed values and understanding of the world. This is analysed using causal connectors indicating justifications, 
and discourse particles signaling speaker stance/attitude; iv) \textit{Persuasiveness} measures deliberative intentions of stakeholders via types of speech acts.
While Gold \textit{et al.} focus on quality criteria that are linguistically externalized considering single arguments, our framework is targeted at revealing structural patterns in the way certain groups argue. 

Bergmann \textit{et al.}~\cite{bergmann} are concerned with providing comprehensive overviews of ongoing debates, to make human decision makers aware of arguments and opinions related to specific topics. 
Their approach relies on a case-based reasoning (CBR) system that allows them to compute similarity between arguments in order to retrieve or cluster similar arguments. CBR also supports the synthesis of new arguments by extrapolating and combining existing arguments. 
Unlike Bergmann \textit{et al.} who focus on grouping or retrieving related arguments, we  propose a data model that focuses less on the analysis and retrieval of single arguments, but aims to provide an aggregate analysis of debates in view of their deliberative quality aspects. 

Bögel \textit{et al.}~\cite{boegel} 
have proposed a rule-based processing framework for analyzing argumentation strategies 
that 
relies on deep linguistic analysis. Their focus is on the 
operationalizaton 
of argument quality that relies on
two central linguistic features: causal discourse connectives and modal particles. The proposed visualization allows users
to zoom into the discourse. However, no aggregate analyses at the level of the whole debate is proposed, as we do in our paper. 

Reed \textit{et al.} have developed several tools to support the exploration and querying of arguments. ACH-Nav~\cite{ZografistouVLR22}, for instance, is a tool for navigating hypotheses that offers access to contradicting hypotheses/arguments for a given hypothesis. Polemicist~\cite{polemicist} %
allows users to explore people's opinions and contributions to the BBC Radio 4 Moral Maze program. ADD-up~\cite{pluss} is an analytical framework that analyzes online debates incrementally, allowing users to follow debates in real time. However, none of these tools are based on a data model that captures the perspectives of different stakeholders in a debate at a structural level. 

VisArgue is an analytical framework by %
Gold \textit{et al.}~\cite{visargue} that focuses on the analysis of debates on a linguistic level, focusing on discourse connectives. A novel glyph-based visualization is described that is used to represent %
instances where similar traits are found among different elements in the dataset. 
More recently, this approach has been extended to analytics of multi-party discourse~\cite{el-assady-etal-2017-interactive}. 
The underlying system combines discourse features derived from shallow text mining with more in-depth, linguistically-motivated annotations from a discourse processing pipeline. Rather than revealing structural patterns in the way different stakeholders argument, the visualisation is designed to give a high-level overview of the content of the transcripts, based on the concept of lexical chaining.

\section{Conclusion}

PAKT, the Perspectivized Argumentation Knowledge Graph and Tool, introduces a pioneering framework for analyzing debates structurally and revealing patterns in argumentation across diverse stakeholders. It employs premises, conclusions, frames, and values to illuminate perspectives, while also enabling the categorization of individuals into socio-demographic groups.

Our application of PAKT to \texttt{debate.org} underscores its efficacy in conducting global analyses and offering valuable insights into argumentative perspectives. In our case studies we demonstrated the versatility of combining perspectivizing categories (\textit{frames, values}) emphasized by different camps, in combination with concept-level analysis -- which enable identification of differences within overall similarities, at the level of individual and across different issues, and how such analyses may indicate starting points for deliberation processes.

PAKT offers broad potential applications by automatically detecting imbalances or underrepresentations in arguments or debates through analyzing frames, values and concepts. Navigation through the PAKT graph via central concepts or argument-similarity edges enhances argument mining to a comprehensive level. This accessible tool allows researchers without a computer science background to explore opinion landscapes at both debate and single-argument levels. Its extensive applications include informing policy-making by dissecting contentious issues and fostering constructive discussions. Integrating PAKT into social media platforms holds promise for highlighting common ground and areas of disagreement among participants, as well as aiding moderators in identifying potentially radical or offensive content. Thus, PAKT serves as a tool to enhance understanding, and also to improve deliberative debates for all. 

\section*{Limitations}

Our analysis and case study rely on automatically annotated data encompassing frames, values, and concepts. Consequently, we anticipate some degree of noise in our dataset, potentially compromising the depth of our analysis. To address this concern, we employ established methodologies derived from prior research to mitigate such discrepancies. Additionally, we perform manual annotations to gauge the quality of our data. 

Our focus lies on the unique aspect of perspectivization, which is not largely explored in prior work. Consequently, we could not directly compare PAKT with other analysis tools from related studies. We hope that our discussion sparks further research, and that PAKT can serve as a valuable baseline in future work. 

Lastly, our analysis and case study shed light on the practical application of PAKT in illuminating insights within debates, thereby aiding in opinion formation and decision-making processes. However, demonstrating PAKT's utility for other tasks such as moderation remains an avenue for future exploration. 

\section*{Acknowledgements}
We thank the anonymous reviewers for their valuable comments, and our annotators, Janosch Gehring and Rahki Asokkumar Subjagouri Nair, for their support. We also thank Marcel Nieveler for his support in the development of the web interface to PAKT. This work has been funded by the DFG through the project ACCEPT as part of the Priority Program “Robust Argumentation Machines” (SPP1999).

\FloatBarrier
\section*{Appendix} 
\FloatBarrier
\begin{figure}
    \centering
    \includegraphics[width=0.72\linewidth]{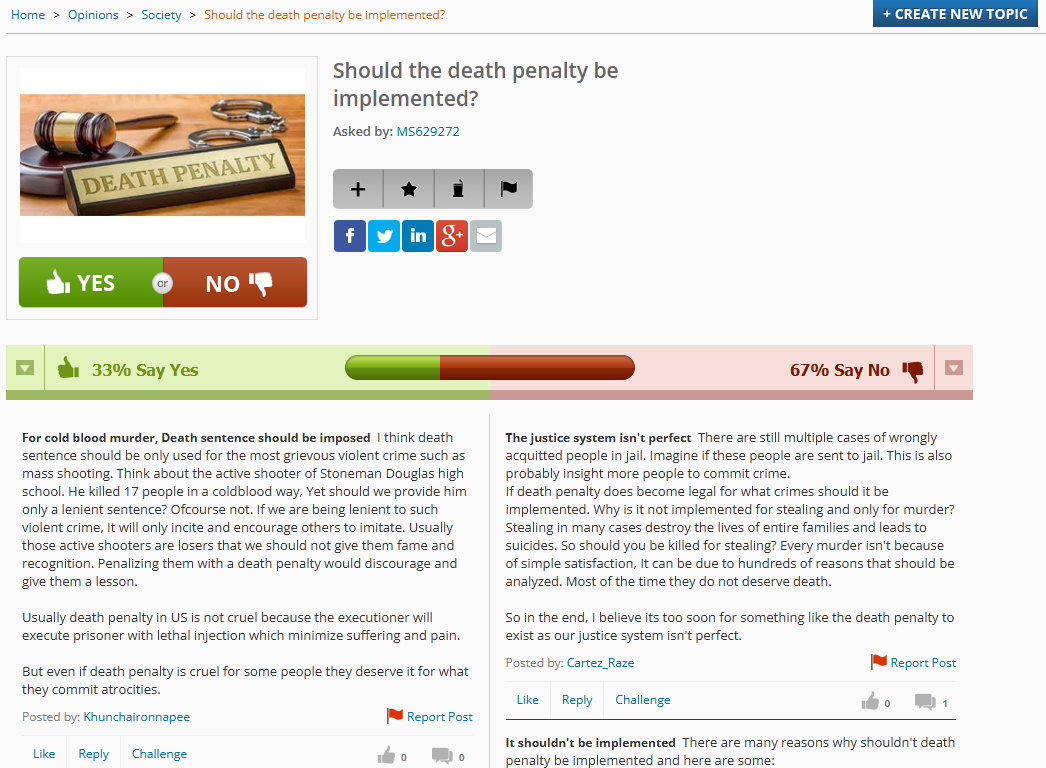}
    \caption{Screenshot of an opinion poll on \texttt{debate.org}}
    \label{fig:debateOrgOpinionPoll}
\end{figure}

\vspace{-1cm}

\FloatBarrier
\bibliographystyle{splncs04}
\bibliography{references}

\section*{Supplementary Materials}
Below are the supplementary materials, which are not part of the peer reviewed main paper~\cite{plenz-etal-2024-pakt}. 

\section{Data source -- DDO} 

The debate portal \texttt{debate.org} (DDO) was a major platform to %
have \textit{one-on-one debates} over several rounds, or to express opinions in the so-called \textit{opinion polls} (see Fig.~\ref{fig:debateOrgOpinionPoll}). The portal was operated from October 2007 to June 2022 and was then closed due to low activity on the portal. However, in the $\sim$15 years of operation, many users debated about various topics, resulting in 81,800 debates (conducted in 433,850 rounds in total), 32,807 opinions, and 53,039 registered in February 2022.

To the best of our knowledge, %
Wachsmuth \textit{et al.}~\cite{wachsmuth:2017f} were the first to crawl DDO in 2017, in order to enrich the argumentative database for the argument search engine \texttt{args.me}. However, the argument search engine contains only 28,045 often verbose one-on-one debates and misses the user profiles and opinion polls. This was changed by a crawl published one year later~\cite{durmus-cardie-2018-exploring,durmus-cardie-2019-corpus} which adds the user profiles and votes of these users rating the one-on-one debates in order to research the relationship between positive ratings and overlapping user attributes.

However, none of the previous works include opinion polls, where users could ask questions, and other users could answer with \textit{yes} or \textit{no}, together with a brief explanation. Therefore, we automatically crawled this 24,646 answers from this part of the portal, shortly before the portal was shutdown. Users could answer anonymously, or they could delete their profile without deleting associated posts. %
Hence, only 7,001 of the 24,646 crawled opinions had a crawlable user profile.

\subsection{Conclusion generation}

For generating the conclusion, we used \texttt{gpt-3.5-turbo} by OpenAI, using the following decoding parameters: with a temperature of $0.5$, we sampled from the entire vocabulary according to the predicted probabilities ($top_p=1.0$). Given the topic $t$ stated as a question, the reply $r$ to the topic, and the stance $s$ (Yes (pro)/No (con)) as given in \texttt{debate.org}, the model was tasked to generate a conclusion as a self-contained statement containing at most $\text{number\_of\_tokens}(t)+\text{number\_of\_tokens}(r)+5$ tokens. Our prompt is the following:

\begin{quote}
    Task: Generate a one-sentence conclusion, given title and reply. Here are a few examples:
    
    Title: Is there anything wrong about homosexuality and SSM? If so, what? (If you comment, please send me a message so we can discuss further.)
    
    Reply: [No] It is morally unethical?
    
    Conclusion Claim: Homosexuality and SSM is morally unethical.
    
    Title: Should presidents be able to use tax money to take vacations during their presidency?
    
    Reply: [Yes] Yes the should
    
    Conclusion Claim: Presidents should be able to use tax money to take vacations during their presidency.
    
    Title: Are unicorns real?
    
    Reply: [No] More real then your brain cells
    
    Conclusion Claim: Intelligence disproves the existence of unicorns.
    
    Please generate the conclusion claim now.
    
    Title: $t$
    
    Reply: [$s$] $r$
    
    Conclusion Claim:    
\end{quote}

\subsection{Frames}

For frame classification, we rely on the annotated data from the MediaFrames-dataset~\cite{card-etal-2015-media}. However, since this dataset contains span annotations for newspaper articles assuming to be framed with exactly one frame class, we have to preprocess and augment the data first. In order to yield a classification model in a multi-class, setting, we present two different model approaches.

\paragraph{Dataset preprocessing} As already stated in the main paper, we considered the 80,146 labeled in-article-text spans on sentence level (in the case of having a text span annotation exceeding the border of a sentence, all sentences are considered which contain at least a part of the annotated text span). In the case of having the votes of multiple annotators, instead of aggregating the frame labels into a majority vote~\cite{HeinischCimiano+2021+59+72}, we combine all votes into a frame target set, containing each frame class that was voted at least once.

In addition, we augment the dataset by merging two consecutive instances by combining their textual representation and unifying their target frameset.

We experiment with two datasets. The \emph{full} dataset consists of all instances, including the augmented ones, resulting in 160,311 instances in total ($\varnothing 1.73$ frames per instance, 48\% of them are assigned with only a single frame label). The \emph{strict} dataset is a subset, excluding instances from the original dataset that were annotated by less than three annotators or are controversial, identified by a low inter-annotator agreement ($1+\frac{1-\# \text{set}(\text{selected frames})}{\# \text{annotators}} < 0.3$). This subset has 18,529 instances in total, and, due to the exclusion of sparsely annotated text spans, we observe a higher density of frame classes ($\varnothing 3.27$ frames per instance, 2\% of them are assigned with only a single frame label).

Each dataset was split into a train, development, and test split with 80\%, 10\%, and 10\% of the dataset, respectively.

\paragraph{Classification models} For the selection of classification models, inspired by Heinisch \textit{et al}.~\cite{heinisch-etal-2023-accept}, we test two approaches: We fine-tune the LLM \texttt{roberta-base}, or we train a shallow neural network (NN), which encodes the token representations calculated by SBERT~\cite{reimers-gurevych-2019-sentence} (using \texttt{all-mpnet-base-v2}). For the NN we use the ELU-activation function, and finally aggregate these processed vectors by averaging them and calculating the standard deviation. A last feed-forward layer computes the final frame probability distribution using these two aggregated vectors. For training the LLM or the shallow neural net, we use early stopping\footnote{with patience of two epochs, measuring the macro-F1-score on the development split} (max. 5 epochs), and a batch size of 16. The learning rate is $5e-5$ and $5e-4$ for the LLM and NN, respectively.

\paragraph{Results} The results are in Table~\ref{tab:app:frame-classification:stats}. We observe the superior performance of LLMs. Looking at our MediaFrames-datasets, using the \emph{strict} data results in an impressive micro-F1 score of 90.2\% on the test split of the \emph{strict} dataset. Using the \emph{full} dataset worsens the F1 to 79.6\% and 73.2\% micro-F1 and macro-F1, respectively. However, looking at the manual annotations on the \texttt{debate.org}-data as described in Section \ref{sec:app:annotation_details}, using the \emph{full} dataset shows better generalization capabilities towards arguments posted in \texttt{debate.org}, yielding 39.7\% and 27.2\% in micro-F1 and macro-F1, respectively. While the \emph{full}-fine-tuned LLM yields a higher precision on \texttt{debate.org}, having the low average of frame classes per instance in mind, the \emph{strict}-fine-tuned LLM is too sensitive in frame detection on \texttt{debate.org}, confirmed by the low precision but relatively high recall. Therefore, we continue our studies with the \texttt{roberta-base}\footnote{published at \url{https://huggingface.co/pheinisch/MediaFrame-Roberta-recall}}, listing more detailed scores in Table \ref{tab:frames}.

\begin{table}[]
    \centering
    \begin{tabular}{|r|cc|cc|}
        \hline
         & \multicolumn{2}{|c|}{Test MediaFrames~\cite{card-etal-2015-media}} & \multicolumn{2}{|c|}{PAKT-annotations}\\
         Setup & micro-F1 & macro-F1 & micro-F1 & macro-F1\\
         \hline
         full data $\mapsto$ RoBERTa & 79.6 & 73.2 & \textbf{39.7} & \textbf{27.2}\\ %
         strict data $\mapsto$ RoBERTa & \textbf{90.2} & \textbf{80.7} & 34.9 & 24.8\\ %
         strict data $\mapsto$ NN & 74.6 & 65.6 & 16.4 & 11.4\\
         \hline
    \end{tabular}
    \caption{Performance of our framing approach (\%) for three selected experimental setups.}
    \label{tab:app:frame-classification:stats}
\end{table}

\subsection{Human Values}

We used the public API at \url{https://values.args.me/api} for requesting the human value detection given an argument. This API relies on the Adam Smith human value detector, which performed best in the ValueEval'23 competition~\cite{schroter-etal-2023-adam,kiesel-etal-2023-semeval}. More precisely, the classification model \textit{EN-Deberta-F1} was used, an ensemble of three models that performed best in the ablation tests.

\subsection{Concepts} \label{sec:construction_concepts}
We adopt Plenz \textit{et al.}~\cite{plenz-etal-2023-similarity} to connect these sentences with ConceptNet subgraphs. To obtain more expressive relations, we remove the unspecific \texttt{RelatedTo} relation. 
First, we use \texttt{en\_core\_web\_trf} from spaCy~\cite{honnibal-etal-2017-spacy} to split premises and conclusions in individual sentences (cf.~\cite{heinisch-etal-2023-accept}), and remove sentences with less than 3 words. 
Then, for each sentence, we use SBERT~\cite{reimers-gurevych-2019-sentence} to identify the most similar triplet in ConceptNet, yielding 2 concepts for each sentence. We connect these concepts by weighted shortest paths. Combining these paths yields a graph for each argument. All concepts along the paths are considered as concepts of the argument.

\subsection{Authors and camps}

Users could provide personal information on DDO. We group certain answers, to obtain a more coarse grained categorization. This enables more analysis, as more data is available for each category. In the following, we list our clustering. In the rare instances where users provide free-form text for certain fields, they are assigned to the "Unknown" category.

\paragraph{\textbf{Ideology}}\ \\
\textbf{Left}: Anarchist, Communist, Green, Liberal, Libertarian, Socialist \\
\textbf{Right}: Conservative, Moderate, Progressive \\
\textbf{Unknown}: Labor, Other, Apathetic, Not Saying, Undecided

\paragraph{\textbf{Income}}\ \\
\textbf{Low}: Less than \$25,000, \$25,000 to \$35,000 \\
\textbf{Medium}: \$35,000 to \$50,000, \$50,000 to \$75,000, \$75,000 to \$100,000 \\
\textbf{High}: \$100,000 to \$150,000, More than \$150,000 \\
\textbf{Unknown}: Not Saying, Other

\paragraph{\textbf{Ethnicity}}\ \\
\textbf{Person of color}: Asian, East Indian, Black, Latino, Other, Middle Eastern, Native American, Pacific Islander \\
\textbf{White}: White \\
\textbf{Unknown}: Not Saying

\paragraph{\textbf{Gender}}\ \\
\textbf{Female}: Female \\
\textbf{Male}: Male \\
\textbf{Diverse}: Genderqueer, Agender, Bigender, Transgender Female, Transgender Male, Androgyne \\
\textbf{Unknown}: Prefer not to say

\paragraph{\textbf{Faith}}\ \\
\textbf{Yes}: Christian, Christian - Methodist, Christian - Protestant, Christian - Lutheran, Christian - Baptist, Christian - Catholic, Christian - Pentecostal, Christian - Latter-Day Saints, Christian - Assemblies of God, Christian - Church of Christ, Christian - Anglican, Christian - Greek Orthodox, Christian - Presbytarian, Christian - Episcopalian, Christian - Seventh-Day Adventist, Christian - Jehovah's Witness, Christian - Amish, Christian - Mennonite, Spiritism, Islamic, Muslim - Sunni, Muslim - Shiite, Muslim - Sufi, Muslim, Yazdânism, Buddhist, Buddhist - Vajrayana, Buddhist - Mahayana, Buddhist - Theravada, Hindu, Hindu - Vaishnavism, Hindu - Saivite, Hindu - Smartha, Hindu - Shakta, Jain, Jewish - Reform, Jewish - Conservative, Jewish - Orthodox, Jewish, Cao Dai, Taoism, Pagan, Neo-Paganism, Mazdakism, Primal-Indigenous, Deism, Unitarian Universalist, Shinto, Scientology, Sikh, Bahá'í, Bábism, Confucian, African Traditional \& Diasporic, Wikkan, Rastafarianism, Zoroastrianism, Yarsani, Mandaeism, Manichaeism, Daoist, Zurvanism, Yazidi, Tenrikyo, Pastafarian, Discordian \\
\textbf{No}: Atheist, Secular, Juche \\
\textbf{Unknown}: Not Saying, Agnostic, Other

\paragraph{\textbf{Education}}\ \\
\textbf{Low}: High School \\
\textbf{Medium}: Some College, Associates Degree \\
\textbf{High}: Bachelors Degree, Graduate Degree, Post Doctoral \\
\textbf{Unknown}: Not Saying, Other

\subsection{Interpretable argument similarity from concept overlap}
To obtain the similarity between two arguments, we consider the Jaccard similarity between their concepts:
\begin{equation}
    S_{Jaccard} = \frac{\left| C_1 \cup C_2 \right|}{\left|C_1 \cap C_2 \right|}, 
\end{equation}
where $C_1$ and $C_2$ are sets consisting of all concepts of argument 1 and 2, respectively. 

Inspired by TF-IDF we also consider weighting by equivalents to term frequency (TF) and inverse document frequency (IDF). As IDF we consider the frequency of the concept occuring in an argument, accross DDO. As TF, we consider the pagerank of the concept in the graphs obtained from ConceptNet, as described in \S\ref{sec:construction_concepts}.

\section{Details to preliminary evaluation}\label{sec:app:annotation_details}

To evaluate the performance of our conclusion generation process, frame identification, human value detection, and approaches measuring the similarity between arguments, we manually labeled 99 arguments on the issue ``Should animal hinting be banned?''. We paid three annotators matriculated in the field of computer science/ computational linguistics. After reading the guidelines and discussing together one example, each annotator labeled the 99 arguments independently from each other.

Considering the evaluation of the \textbf{generated conclusion}, we ask ``Is the AI-generated conclusion appropriate? (reflects the stance regarding the topic, incorporates the argumentative text)''. Considering the majority vote, 94.9\% of the conclusions were appropriate. Having a more fine-grained analysis, 40.4\% of the conclusions were labeled as (very) good, having a conclusion representing the correct stance and also being very tailored to the presented premise. 28.3\% of the conclusions have the correct stance, but are very generic, e.g. ``Animal hunting shouldn't be banned.'' 26.3\% of the conclusions are incomplete in the sense of hiding the central concluding points of the premise. Only 5.1\% of all conclusions generated by ChatGPT were inappropriate. These conclusions do not reflect or even contrast the central point made by the premise. None of the presented conclusions convey the opposite stance or were linguistically broken. The inter-annotator-agreement is moderate (Fleiss' kappa $\kappa = 0.492$).

Considering the evaluation of the classified \textbf{frames}, we ask ``Which frames (=aspects) are emphasized by the argument?'' without showing the predicted classes. We design this task as a multiple-choice task. Table~\ref{tab:frames} presents the results, distinguishing between selecting all frame classes as ground truth if they are manually selected at least by one annotator (one-hit vote), as well as the majority vote, and the full agreement (only accepting a frame class as ground truth if all annotators identify this class). Considering the one-hit vote, the observed micro-F1 score is 32.7 ($\varnothing 4.6$ frame classes per argument). For the majority vote, we yield a micro-F1 score of 39.7 ($\varnothing 2.1$ frame classes per argument), and for full agreement 37.2 ($\varnothing 1.0$ frame classes per argument). The prediction performance differs between the single frame classes. While an economic framing is easy to detect (mainly looking for concurrency symbols as ``\$''), ``fairness and equality'' and ``public opinion'' are often more implicitly expressed. Even for our annotators, it was not easy to decide when a public opinion is mentioned or a fallacy of composition. The overall inter-annotator-agreement is moderate (Fleiss' kappa $\kappa = 0.213$), which is comparable to other manual studies about frame identification~\cite{card-etal-2015-media}. However, the agreement largely differs between the frame classes. For example, the ``economic'' frame ($\kappa = 0.62$) is not as controversial as the frame ``public opinion'' which was rarely observed by two annotators where the third annotator has a plausible reasoning for not selecting this frame due to its weak focus.

\begin{table}[]
    \centering
    \begin{tabular}{|r|c|ccc|c|}
        \hline
        & One-hit vote & \multicolumn{3}{c|}{Majority vote} & Full agreement\\
        Frame & F1 & Precision & Recall & F1 & F1 \\
        \hline
        economic & 44.4 (14) & \textbf{100.0} & 57.1 & \textbf{72.7 (7)} & 66.7 (5)\\
        capacity \& resources & 28.9 (71) & 83.3 & 22.2 & 35.1 (45) & 37.0 (15)\\
        morality & 53.7 (79) & 89.7 & 39.4 & 54.7 (66) & 55.7 (50)\\
        fairness \& equality & 16.7 (64) &  75.0 & 17.6 & 28.6 (34) & 30.0 (12)\\
        legality & 7.1 (26) & 50.0 & 33.3 & 40.0 (3) & 0.0 (0)\\
        policy prescription & 25.6 (16) & 0.0 & 0.0 & 0.0 (0) & 0.0 (0)\\
        crime \& punishment & 16.7 (7) & 20.0 & \textbf{100.0} & 33.3 (1) & 0.0 (0)\\
        security \& defense & 0 (21) & 0.0 & 0.0 & 0.0 (3) & 0.0 (1)\\
        health \& safety & \textbf{55.4 (23)} & 21.4 & 81.8 & 34.0 (11) & 4.7 (1)\\
        quality of life & 37.3 (58) & 52.9 & 36.0 & 42.9 (25) & 58.1 (14)\\
        cultural identity & 30.4 (39) & 85.7 & 54.5 & 66.7 (11) & \textbf{72.7 (4)}\\
        public opinion & 11.1 (34) & 0.0 & 0.0 & 0.0 (6) & 0.0 (0)\\
        political & 0.0 (1) & 0.0 & 0.0 & 0.0 (0) & 0.0 (0)\\
        external regulation & 0.0 (1) & 0.0 & 0.0 & 0.0 (0) & 0.0 (0)\\
        other & 0.0 (1) & 0.0 & 0.0 & 0.0 (0) & 0.0 (0)\\
        \hline
        \hline
        micro avg & 32.7 (454) & 47.7 & 34.0 & 39.7 (212) & 37.2 (102)\\
        macro avg & 21.8 (454) & 38.5 & 29.5 & 27.2 (212) & 21.7 (102)\\
        \hline
    \end{tabular}
    \caption{Performance of our framing approach (\%).
    The numbers in brackets represent the total number of arguments labeled with the responding frame class.}
    \label{tab:frames}
\end{table}

Considering the evaluation of the classified \textbf{human values}, we ask ``Which human values play a role in/ impress this argument?'' without showing the predicted classes. We design this task as a multiple-choice task. Table~\ref{tab:values} presents the results in a similar fashion to Table~\ref{tab:frames} about frame identification. Similar to frame identification, the occurrences as well as the prediction performances are not equally distributed among the single value classes. Considering the majority vote label aggregation, the value ``hedonism'' was always detected by our automatic approach, yielding an F1-score of 76.9\%. Other values such as ``humility'' were more implicitly expressed using fewer trigger words, and often debatable in the annotation process, yielding an F1-score of only 16.9\%.
We observe only a slight agreement of $\kappa=0.149$ in this task. Besides the subjective nature of the task requiring looking ``between the lines'' and selecting a subset of 20 classes, some values were not sufficiently defined for the issue of animal hunting. For example, there was the question of whether the value ``benevolence: caring'' applies to animals as well, pointing to the margin of interpretation to the definition\footnote{following \url{https://touche.webis.de/semeval23/touche23-web/\#task}} and the appliance of these human values.

\begin{table}[]
    \centering
    \begin{tabular}{|r|c|ccc|c|}
        \hline
        & One-hit vote & \multicolumn{3}{c|}{Majority vote} & Full agreement\\
        Human Values & F1 & Precision & Recall & F1 & F1 \\
        \hline
        self-direction: thought & 0.0 (4) & 0.0 & 0.0 & 0.0 (0) & 0.0 (0)\\
        self-direction: action & 42.3 (17) & 17.1 & \textbf{100.0} & 29.3 (6) & 5.6 (1)\\
        stimulation & 47.6 (14) & 42.9 & 60.0 & 50.0 (5) & 22.2 (2)\\
        hedonism & 58.8 (9) & \textbf{62.5} & \textbf{100.0} & \textbf{76.9 (5)} & \textbf{40.0 (2)}\\
        achievement & 32.3 (16) & 13.1 & 50.0 & 21.1 (4) & 11.8 (2)\\
        power: dominance & 0.0 (14) & 0.0 & 0.0 & 0.0 (6) & 0.0 (3)\\
        power: resources & 18.2 (40) & 25.0 & 12.5 & 16.7 (8) & 33.3 (2)\\
        face & 0.0 (4) & 0.0 & 0.0 & 0.0 (0) & 0.0 (0)\\
        security: personal & 66.7 (46) & 42.9 & 93.6 & 58.8 (16) & 20.5 (4)\\
        security: societal & 23.1 (13) & 0.0 & 0.0 & 0.0 (2) & 0.0 (0)\\
        tradition & \textbf{70.8 (29)} & 52.6 & 83.3 & 64.5 (12) & \textbf{40.0 (6)}\\
        conformity: rules & 50.8 (31) & 6.2 & 66.7 & 11.4 (3) & 0.0 (0)\\
        conformity: interpersonal & 0.0 (5) & 0.0 & 0.0 & 0.0 (0) & 0.0 (0)\\
        humility & 7.3 (51) & 50.0 & 9.5 & 16.9 (21) & 0.0 (3)\\
        benevolence: caring & 43.0 (54) & 28.9 & 50.0 & 35.9 (14) & 14.3 (3)\\
        benevolence: dependability & 30.8 (8) & 0.0 & 0.0 & 0.0 (0) & 0.0 (0)\\
        universalism: concern & 62.0 (29) & 11.9 & 62.5 & 20.0 (8) & 0.0 (0)\\
        universalism: nature & 85.7 (65) & 48.8 & 97.6 & 65.0 (41) & 37.6 (19)\\
        universalism: tolerance & 16.7 (11) & 0.0 & 0.0 & 0.0 (1) & 0.0 (0)\\
        universalism: objectivity & 51.2 (22) & 23.8 & 83.3 & 37.0 (6) & 16.7 (3)\\
        \hline
        \hline
        micro avg & 50.6 (481) & 28.5 & 65.2 & 39.7 (158) & 18.5 (50)\\
        macro avg & 35.4 (481) & 21.3 & 43.5 & 25.1 (158) & 12.1 (50)\\
        \hline
    \end{tabular}
    \caption{Performance of our value-detection approach (\%).
    The numbers in brackets represent the total number of arguments labeled with the responding frame class.}
    \label{tab:values}
\end{table}

Considering the automatically calculated \textbf{similarities between arguments}, we propose two tasks. One task was, given the presented main argument and two other arguments of the same topic and stance, rating whether and which of the two other arguments are more similar to the main argument. The other task aims to relate similarities with counter-arguments. Hence, given the presented main argument and two other arguments of the same topic \emph{but opposite stance}, rating whether and which of the two other arguments counters or attacks the main argument more. Hence, we labeled the similarity in a relative fashion while having absolute predictions. To compare the human labels with the automatically calculated labels, we map these absolute predictions into the relative setting of the annotation study by pairing the predicted similarity numbers with $(sim_1, sim_1) = (sim(a, a_1), sim(a, a_2))$, having $a$ as the main argument and $a_1$ as the first other argument and $a_2$ as the second one.

\begin{equation}
    \text{predicted label} = \left\{
    \begin{array}{ll}
        a_1 \text{ is more similar/ counter} & \quad sim_1-sim_2 > \theta \\
        a_2 \text{ is more similar/ counter} & \quad sim_2-sim_1 > \theta \\
        \textrm{equal} & \quad \text{else}\\
    \end{array}
    \right.
    \label{eq:map_sim}
\end{equation}

We finally apply Equation \ref{eq:map_sim} to the pairs of predicted similarities, setting a threshold $\theta$ for predicting ``equal'' as a hyperparameter. Table~\ref{tab:simcount} shows the results for $\theta \in \{0, 0.1, 0.33\}$. We observe the best macro-F1-values with $\theta=0.1$, representing the instruction to the annotators to not overuse that class but only label such argument pairs as ``equal'' if there are no tendencies for one side regarding similarity (19/99 cases) or counter (25/99 cases) are observable. Regarding similarity prediction, the global S$^3$BERT is outperforming with a macro-F1 score of 42.7\% (42\% accurate classifications) with $\theta = 0.1$. Regarding best counterargument-prediction, S$^3$BERT considering negations in the arguments is outperforming, yielding 40.4 macro-F1 (40\% accurate classifications) with $\theta = 0.1$. Predicting the similarity by calculating similarities between the concept sets of both arguments has a lower performance but is as a symbolic approach more explainable as S$^3$BERT.
The low performances indicate the difficulty of the task of considering relational argument similarity and how well arguments counter each other -- in addition to its subjective component. The task of similarity and counterarguments, we observe a fair agreement of $\kappa=0.251$ and a slight agreement of $\kappa=0.105$, respectively.

\begin{table}[]
    \centering
    \begin{tabular}{|l|cc|cc|cc|}
        \hline
        & \multicolumn{2}{c|}{$\theta=0.0$} & \multicolumn{2}{c|}{$\theta=0.1$} & \multicolumn{2}{c|}{$\theta=0.33$}\\
        Approach & Sim & Cou & Sim & Cou & Sim & Cou\\
        \hline
        S$^3$BERT & 38.6 & 33.2 & \textbf{42.7} & 35.0 & 15.3 & 18.3\\
        S$^3$BERT -- concepts & 34.3 & \textbf{35.4} & 35.6 & 38.8 & 19.1 & 17.2\\
        S$^3$BERT -- negations & 31.3 & 32.4 & 35.4 & \textbf{40.4} & \textbf{20.1} & \textbf{23.9}\\
        \hline
        Concepts -- Jaccard & 37.8 & 30.2 & 27.1 & 26.0 & 12.5 & 13.4 \\
        Concepts -- IDF     & \textbf{40.2} & 34.2 & 16.9 & 14.8 & 10.7 & 13.4 \\
        Concepts -- TF-IDF   & 38.7 & 34.1 & 24.3 & 23.3 & 15.8 & 13.4 \\
        \hline
    \end{tabular}
    \caption{Macro-F1 (\%) of argument-similarity predictions, once for the similarity annotation task (sim) and once for the counter annotation task (cou).}
    \label{tab:simcount}
\end{table}

In the last step, we evaluate a part of metadata from \texttt{debate.org}, specifically the stance of an argument. The task reads as ``According to the presented argument, should animal hunting be banned?''. As expected, nearly all stance labels were confirmed by our annotators. Only 2/99 arguments have a doubtful stance label. For this task, we observe an almost perfect agreement of $\kappa=0.973$.

\section{Details to frames and human values}

\subsection{Description of the used frame and value classes}

In this Section, we list and briefly describe all frame classes and used human values categories. The descriptions follow the annotation guidelines for measuring the quality of our PAKT-data.

\subsubsection{MediaFrames by Card \textit{et al}.~\cite{card-etal-2015-media}}

\begin{enumerate}
    \item ECONOMIC: costs, benefits, or other financial implications
    \item CAPACITY AND RESOURCES: availability of physical, human or financial resources, and capacity of current systems
    \item MORALITY: religious or ethical implications
    \item FAIRNESS AND EQUALITY: balance or distribution of rights, responsibilities, and resources
    \item LEGALITY, CONSTITUTIONALITY AND JURISPRUDENCE: rights, freedoms, and authority of individuals, corporations, and government
    \item POLICY PRESCRIPTION AND EVALUATION: discussion of specific policies aimed at addressing problems
    \item CRIME AND PUNISHMENT: effectiveness and implications of laws and their enforcement
    \item SECURITY AND DEFENSE: threats to the welfare of the individual, community, or nation
    \item HEALTH AND SAFETY: health care, sanitation, public safety
    \item QUALITY OF LIFE: threats and opportunities for the individual’s wealth, happiness, and well-being
    \item CULTURAL IDENTITY: traditions, customs, or values of a social group in relation to a policy issue
    \item PUBLIC OPINION: attitudes and opinions of the general public, including polling and demographics
    \item POLITICAL: considerations related to politics and politicians, including lobbying, elections, and attempts to sway voters
    \item EXTERNAL REGULATION AND REPUTATION: international reputation or foreign policy
    \item  OTHER: any coherent group of frames not covered by the above categories
\end{enumerate}

\subsubsection{Human Values by Schwarzt \textit{et al}.~\cite{schwartz-1994-universal}}

The selected granularity level of human values and their descriptions base upon the shared task about human values by Kiesel \textit{et al}.~\cite{kiesel-etal-2023-semeval}.

\begin{enumerate}
    \item \underline{SELF-DIRECTION} -- THOUGHT: encouragement of the individual’s ideas and interests – creative, curious, freedom of thoughts
    \item \underline{SELF-DIRECTION} -- ACTION: encouragement of the individual’s actions, emphasizing choosing own goals, independence, privacy – being self-determined. Everyone should do what [s]he wants
    \item \underline{STIMULATION}: encouragement to experience excitement, novelty, and change of individuals. Call for having an exciting/ varied life (e.g. travel, sport, activities...), risk something
    \item \underline{HEDONISM}: encouragement to experience pleasure and sensual gratification –- enjoy life, emphasizes fun
    \item \underline{ACHIEVEMENT}: encouragement to be successful in accordance with social norms, emphasizing success, intellectual. Aligns with the statement: ``no pain, no gain''
    \item \underline{POWER} -- DOMINANCE: encouragement to be in positions of control over others –- emphasizes hierarchies and aims for influence and the right to command
    \item \underline{POWER} -- RESOURCES: encouragement to have material possessions and social resources. The value occurs in arguments towards allowing people to gain wealth and material possessions.
    \item \underline{FACE}: encouragement to maintain one's public image,  gaining respect. Arguments with this human value emphasize the importance of a  good reputation and point to protection of the public image
    \item \underline{SECURITY} -- PERSONAL: encouragement of the individual’s secure environment, raising the importance of having a sense of belonging (group-caring), emphasizing good health, having no depts, being neat and tidy and aiming for a stable comfortable life
    \item \underline{SECURITY} -- SOCIETAL: encouragement of secure and stable overall society, argues for having a safe country\& stable society
    \item \underline{TRADITION}: encouragement to maintain cultural, family, or religious traditions, argues for respecting traditions, and being based on religious faith/ mission
    \item \underline{CONFORMITY} -- RULES: encouragement to comply with rules, laws, and formal obligations. This human value encourages good manners and being self-disciplined
    \item \underline{CONFORMITY} -- INTERPERSONAL: encouragement to avoid upsetting or harming others, argues for politeness, and honoring elders
    \item \underline{HUMILITY}: encouragement to recognize one's own insignificance in the larger scheme of things. This human value asks for life acceptance as is it, and encourages being satisfied.
    \item \underline{BENEVOLENCE} -- CARING: encouragement to work for the welfare of one's group's members, emphasizing helpfulness/ honesty/ forgiveness/ love, also arguments for having, protecting, and caring  for their family
    \item \underline{BENEVOLENCE} -- DEPENDABILITY: encouragement to be a reliable and trustworthy member of one's group, emphasizing responsibility/ loyalty
    \item \underline{UNIVERSALISM} -- CONCERN: encouragement to strive for equality, justice, and protection for \emph{all} people, aiming a world at peace
    \item \underline{UNIVERSALISM} -- NATURE: encouragement to preserve the natural environment, being in harmony with nature/ striving for a world of natural beauty
    \item \underline{UNIVERSALISM} -- TOLERANCE: encouragement to accept and try to understand those who are different from oneself
    \item \underline{UNIVERSALISM} -- OBJECTIVITY: encouragement to search for the truth and think rationally and in an unbiased way. This human value encourages logical thinking which is not guided by feelings
\end{enumerate}

\subsection{Exemplary error analysis of frame classification and human value detection}

In this Section, we list a few examples where the prediction of frames (Table~\ref{tab:frames:errors}) and human values (Table~\ref{tab:human_values:errors}) diverges from the majority vote collected in the preliminary evaluation (Section~\ref{sec:app:annotation_details}). These examples show us that often the prediction ``errors'' are only a matter of perspective or sensitivity, and are not ``totally wrong''. Our frame classifier tends to output too few frame classes while the human value detector tends to be too verbose.

\begin{table}[]
    \centering
    \begin{tabular}{|p{0.66\linewidth}|p{0.15\linewidth}|p{0.15\linewidth}|}
        \hline
        Argument & Ground truth & Predicted \\
        \hline
        (No human can still support hunting.) The wildlife on our world is already on the brink of disappearing, And yet some inferior minds feel the need to literally chase them to death. You hear many fox hunters say, ''but it's a source of food''. Yeah, Maybe after emptying your freezer full of supermarket meats you can get around to eating that horrible dog meat & \emph{CAPACITY AND RESOURCES} & \textcolor{green}{capacity and resources}, \textcolor{red}{health and safety}\\
        \hline
        (Hunting helps the economy) If the US hunters just stopped hunting we would have an extra 14.98 billion dollars added to the national debt. Also in 2011 the 13.7 million hunters in the US generated 11.8 billion dollars in taxes and spent 38.3 billion to get ready for their own hunting trip. Hey be stupid get rid of hunting see how much more you have to pay in taxes. & \emph{ECONOMIC}, CAPACITY AND RESOURCES & \textcolor{green}{economic}\\
        \hline
        (Please Respect life!) God made us ruler over the earth and over all the creatures that move along the ground. We must only hunt or kill   animal Primarily   for food (or in self-defense) killing an animal for money, trophy or for their skin is really absurd! A life is never worth it as a sacrifice for such things. As a steward we must do something for their conservation not to their extinction! & \emph{CAPACITY AND RESOURCES}, \emph{MORALITY} & \textcolor{green}{morality}\\
        \hline
    \end{tabular}
    \caption{Table showing a frame classification analysis showcasing three examples. \emph{Emphasised printed} classes in the ground truth are voted with a full agreement.}
    \label{tab:frames:errors}
\end{table}

\begin{table}[]
    \centering
    \begin{tabular}{|p{0.66\linewidth}|p{0.15\linewidth}|p{0.15\linewidth}|}
        \hline
        Argument & Ground truth & Predicted\\
        \hline
        (No human can still support hunting.) The wildlife on our world is already on the brink of disappearing, And yet some inferior minds feel the need to literally chase them to death. You hear many fox hunters say, ''but it's a source of food''. Yeah, Maybe after emptying your freezer full of supermarket meats you can get around to eating that horrible dog meat & HUMILITY, \emph{UNIVERSALISM -- NATURE} & \textcolor{red}{universalism: concern}, \textcolor{green}{Universalism: nature}\\
        \hline
        (Hunting helps the economy) If the US hunters just stopped hunting we would have an extra 14.98 billion dollars added to the national debt. Also in 2011 the 13.7 million hunters in the US generated 11.8 billion dollars in taxes and spent 38.3 billion to get ready for their own hunting trip. Hey be stupid get rid of hunting see how much more you have to pay in taxes. & \emph{POWER -- RESOURCES} & \textcolor{red}{Achievement}, \textcolor{green}{Power: resources}, \textcolor{orange}{Security: personal}, \textcolor{orange}{Security: societal}\\
        \hline
        (Please Respect life!) God made us ruler over the earth and over all the creatures that move along the ground. We must only hunt or kill   animal Primarily   for food (or in self-defense) killing an animal for money, trophy or for their skin is really absurd! A life is never worth it as a sacrifice for such things. As a steward we must do something for their conservation not to their extinction! & \emph{UNIVER-SALISM -- NATURE} &  \textcolor{orange}{Tradition}, \textcolor{red}{Conformity: rules}, \textcolor{red}{Benevolence: dependability}, \textcolor{orange}{Universalism: concern}, \textcolor{green}{Universalism: nature}\\
        \hline
    \end{tabular}
    \caption{Table showing a human value detection analysis showcasing three examples. \emph{Emphasised printed} classes in the ground truth are voted with a full agreement. Orange colored classes in the predictions are supported by one annotator (minority), red colored classes in the predictions are never voted.}
    \label{tab:human_values:errors}
\end{table}

\end{document}